\documentclass{amsart}
\usepackage{amsmath,amsfonts}

\usepackage{amscd,amsthm}
\usepackage{graphicx}
\usepackage{chemarrow}
\usepackage{MnSymbol}
\theoremstyle{plain} %% This is the default

\newtheorem{theorem}{Theorem}[section]

\newtheorem{proposition}[theorem]{Proposition}

\newtheorem{remark}[theorem]{Remark}

\usepackage{color}

\newcommand{\documentdate}{6 12 2012}
\usepackage{a4wide,graphicx,color}

\begin{document}

\addtolength{\hoffset}{-1 cm}
\addtolength{\textwidth}{1 cm}
\addtolength{\voffset}{0.2 cm}
\addtolength{\textheight}{0.5 cm}

% ... overwrite A4 top margin to make it readable on letter.
\topmargin -10truept
\pagestyle{myheadings}
\markright{Name and title}

%\title{The title}
%\author{T. Carletti and A. Filisetti}
\date{\documentdate}

%\begin{document}
\begin{titlepage}

\includegraphics[height=3.5cm]{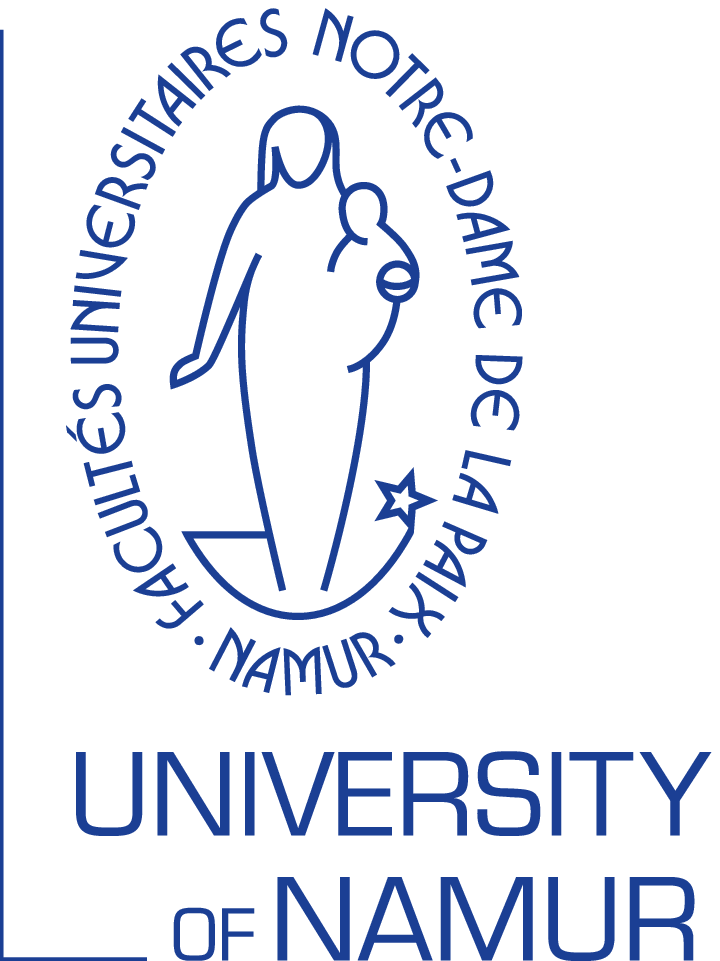}

\vspace*{2cm}
\hspace*{1.3cm}
\fbox{\rule[-3cm]{0cm}{6cm}\begin{minipage}[c]{12cm}
\begin{center}
The stochastic evolution of a
  protocell. \\The 
  Gillespie  algorithm in a 
  dynamically varying volume.
\mbox{}\\
by T. Carletti and A. Filisetti \\
\mbox{}\\
Report naXys-22-2011 \hspace*{20mm} \documentdate 
\end{center}
\end{minipage}
}

\vspace{2cm}
\begin{center}
\includegraphics[height=3.5cm]{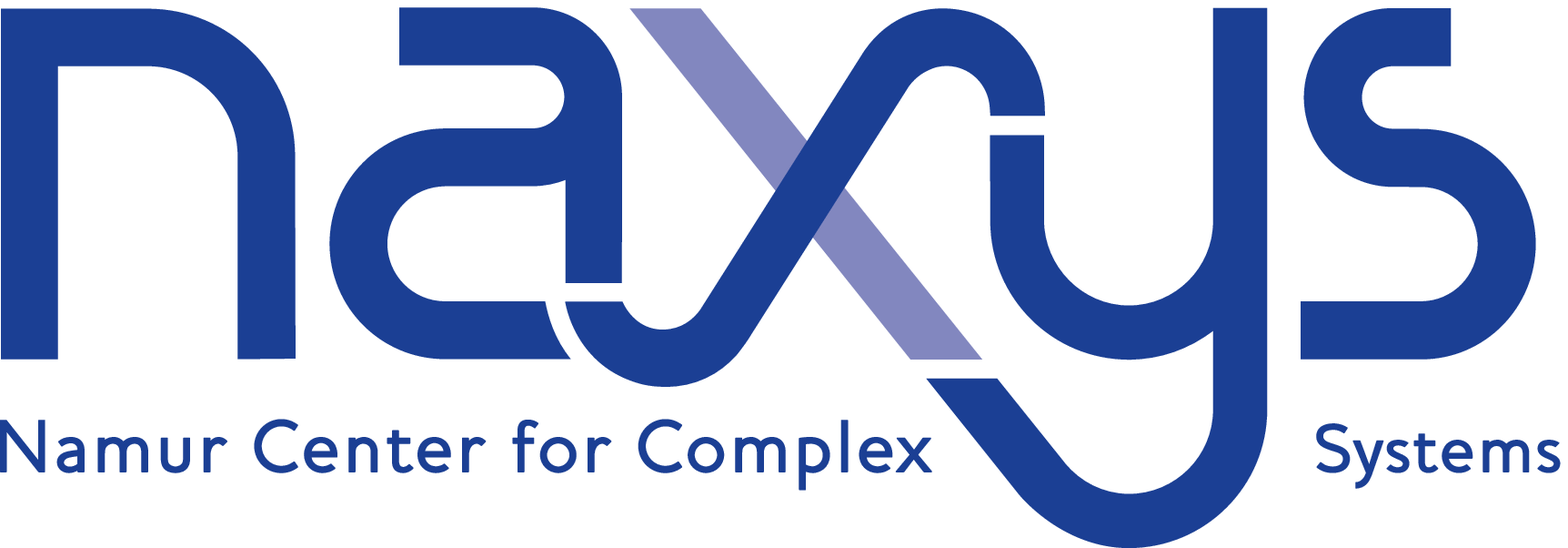}

\vspace{2cm}
{\Large \bf Namur Center for Complex Systems}

{\large
University of Namur\\
8, rempart de la vierge, B5000 Namur (Belgium)\\*[2ex]
{\tt http://www.naxys.be}}

\end{center}

\end{titlepage}

\author{T. Carletti and A. Filisetti}

\address[Timoteo Carletti]{naXys, Namur Center for Complex Systems and
  University of Namur,  Namur, Belgium 5000} 

\address[Alessandro Filisetti]{CIRI, Energy and Environment Interdipartimental Center for Industrial Research, University of Bologna and European Centre for Living Technology, University C\'a Foscari of Venice}
 
\date{\today}
\title[Stochastic evolution of a protocell.]{The stochastic evolution of a
  protocell. \\The 
  Gillespie  algorithm in a 
  dynamically varying volume.}
\maketitle 

\begin{abstract}
In the present paper we propose an improvement of the Gillespie algorithm
allowing us to study the time evolution of an ensemble of chemical reactions
occurring in a varying volume, whose growth is directly related to the amount
of some specific molecules, belonging to the reactions set.

This allows us to study the stochastic evolution of a protocell, whose volume
increases because of the production of container molecules. Several protocells
models are considered and compared with the deterministic models.
\end{abstract}

\section{Introduction}
\label{sec:intro}

All {known} life forms are composed {of} basic units {called} {\em
  cells};  
this holds true from the single-cell prokaryote bacterium to the highly
sophisticated eucaryotes, whose existence is the result of the coordination,
{in term of self-organization and emergence,} of the behavior of each single
basic unit.

While present day cells are endowed with highly sophisticated regulatory
mechanisms, that represent the outcome of almost four billion-years of
evolution, it is generally believed that the first life-forms were much
simpler. Such primordial life-bricks, the {\em protocells}, were most probably
exhibiting only few simplified functionalities, that required a primitive
embodiment structure, a protometabolism and a rudimentary genetics, so to
guarantee that offsprings were \lq\lq similar\rq\rq~to their parents~\cite{Albert,Rasmussen,Szostak}.

Intense research programs are being
established aiming at obtaining protocells capable of growth and duplication,
endowed with some limited form of
genetics~\cite{Mansy2008,Oberholzer1995,Rasmussen2004,Szostak}. Despite all
efforts, artificial protocells have not yet been reproduced in 
laboratory and it is thus extremely important to develop reference
models~\cite{CarlettiJTB,Luisi2006,Rasmussen2004,SerraAlife} 
that capture the essence of the first protocells appeared on Earth and enable
to monitor their subsequent evolution. Due to the uncertainties about the
details, high-level abstract models are particularly relevant. Quoting
Kaneko~\cite{Kaneko2006}  
it is necessary to consider \lq\lq simplified models able to capture universal
behaviors, without carefully adding complicating details\rq\rq .

Most of the models present in the literature are based on {deterministic
  differential equations} governing the evolution of the concentrations of the
involved  {reacting} molecules. Even if the results are worth discussing and
provide important insights, it should be stressed that the former assumptions
are rarely satisfied in a cell~\cite{Gillespie1997}. Firstly, the number of
involved molecules is 
small and should be counted by integer numbers, hence the use of the
concentrations can be questioned; secondly, the presence of the thermal noise
introduces in the system a degree of stochasticity than cannot be trivially
encoded by a differential equation, mostly because this makes the time
evolution a stochastic process. One possible way to overcome such
difficulties is to use the Chemical Master equation: given the present state
of the system, namely the number of available molecules  {for
  each species}, and the possible
reactions  {among them}, one can compute the transition probabilities to reach and leave the given state and thus get a 
partial differential equation describing the time evolution of the probability
distribution of having a given number of molecules at any future
times~\cite{Gillespie1997,GillespieBook}. 

Analytically solving the 
resulting equation is normally a very hard task, one should thus resort to use
numerical 
methods. A particularly suitable one is the algorithm presented by
Gillespie~\cite{Gillespie1997,GillespieBook}, allowing to determine, as a
function of the present state of the system, the most probable reaction and
the most probable reaction time, i.e. the time at which such reaction will
occur. 

Let us however observe that in the setting  {we are hereby
  interested in}, the chemical reactions
occur in a varying volume, because of the 
protocell growth; we thus need to adapt the Gillespie method to
account for this factor. To the best of our knowledge, there are in the
literature very few papers dealing with the Gillespie algorithm in a varying
volume~\cite{kierzek2002,LVTH2004}. Moreover in all these
papers, the volume variation can be considered as an exogenous factor, not
being directly related to the number of lipids {forming} the protocell
membrane. So our main contribution is to improve the Gillespie {algorithm}
taking into account the protocell varying 
volume which is moreover consistent with the increase of the number of lipids
 {constituting the protocell membrane}.

The paper is organized as follow. In Section~\ref{sec:srm} we briefly recall
the Surface Reaction Models of protocell, that would be used to compare our
stochastic numerical scheme. Then in Sections~\ref{sec:meth}
and~\ref{sec:algo} we will 
present our implementation of the Gillespie algorithm in a dynamically varying
volume. Finally in Section~\ref{sec:app} we will present some applications of
our method.

\section{Surface Reaction Models}
\label{sec:srm}

Among the available models for protocells, a particularly interesting one is
the Surface Reactions Model~\cite{CarlettiJTB,SerraAlife}, SRM for short, and
its applicability to the synchronization problem. Such
model is roughly inspired by the Los Alamos bug
hypothesis~\cite{Rasmussen2004,Rasmussen} but which, due to its abstraction
level, the SRM can be applied to a wider set of protocell hypotheses. 

The SRM 
is build on the assumption that a protocell should comprise at least one kind
of \lq\lq container\rq\rq~molecule (typically a lipid or amphiphile), hereby
called $C$ molecule, and one kind of replicator molecule - loosely speaking
\lq\lq genetic material\rq\rq , 
hereafter called Genetic Memory Molecule, GMM for short, and named with the
letter $X$. There are therefore
two kinds of reactions which are crucial for the working of the protocell:
those which synthesize the container molecules Eq.~\eqref{eq:Xi2C} and
those which synthesize the GMM replicators Eq.~\eqref{eq:XiPj2Xj}
\begin{equation}
  \label{eq:Xi2C}
  X_i + L_i \autorightarrow{$\alpha_i$}{}X_i+C \, ,
\end{equation}

\begin{equation}
  \label{eq:XiPj2Xj}
  X_i + P_j \autorightarrow{$M_{ij}$}{}X_i+X_j \, .
\end{equation}
In both cases $L_i$ and $P_j$ are the buffered precursors, respectively of
container molecules and of the $j$--th GMM, while $\alpha_i$ and $M_{ij}$ are the
reaction {kinetic} constants.

A second main assumption of the SRM, is that such reactions occur on the
{\em surface} of the protocell, exposed to the external medium where
precursors are 
free to move. Hence, as long as container molecules are produced, they are
incorporated in the 
membrane that thus increases its size, until a critical point at which, due to
physical instabilities, the membrane splits and two offsprings are obtained,
each one getting half of the mother's GMMs and whose size is roughly half that of the mother just before the division.

Under the previous assumptions and in the deterministic setting, one can
prove~\cite{CarlettiJTB,SerraAlife} 
that the number of membrane molecules and the number
of GMMs evolve in time according to: 
\begin{equation}
  \label{eq:ode}
  \begin{cases}
    \dot{C}&=\left(\frac{C}{\rho}\right)^{\beta-1}\vec{\alpha}\cdot \vec{X}\\
    \dot{\vec{X}}&=C^{\beta-1}M\cdot \vec{X}\, ,
  \end{cases}
\end{equation}
where $\vec{X}=(X_1,\dots,X_N)$ represents the amount of each GMM,
$\vec{\alpha}=(\alpha_1,\dots,\alpha_N)$ is the vector of the reaction
constants responsible for the production of $C$ molecules from the $X$
molecules plus some appropriate precursor. $\left(M_{ij}\right)$ denotes the reaction constant at which $X_i$
is produced by $X_j$ plus some precursor. $\beta\in[2/3,1]$ is a geometrical
shape factor that relates the 
surface to the volume of the protocell and $\rho$ is the lipid density (for
more details the interested reader can
consult~\cite{CarlettiJTB,SerraAlife}).  {Let us observe that in this
  setting the precursors are assumed to be buffered and thus their amount to
  be constant, hence the latter can be incorporated into the constants
  $\alpha$ and $M$.}

So starting with a initial value of container molecules, $C(t_0)=C_0$, and of
GMMs, $\vec{X}(t_0)=\vec{X}_0$, the protocell will grow until some time
$t_0+\Delta T_1$ at which the amount of $C$ molecules has doubled with respect
to the initial value,
$C(t_0+\Delta T_1)=2C_0$ and thus the protocell undergoes a
division. Each offspring will get half of the GMMs the mother protocell had
just before the division, $\vec{X}^{(1)}=\vec{X}(t_0+\Delta T_1)/2$. And the
protocell cycle starts once again.  {One can
  prove~\cite{CarlettiJTB,SerraAlife} that under suitable conditions
  $\vec{X}^{(n)}$ tends to a constant value once $n$ goes to infinity,
  implying thus the emergence of synchronization of growth and information
  production.} 

\section{The method }
\label{sec:meth}

Let us now improve the previous scheme by introducing a probabilistic
setting {\em \`a la Gillespie}. We thus consider a protocell made by a lipidic
vesicle and containing 
a well stirred mixture of $N$ GMMs, $X_1,\dots, X_N$,
that may react through $m$ elementary
reaction channels $R_{\mu}$, $\mu=1,\dots,m$,  running within {the}
volume $V(t)$  {of the protocell}.

Let us observe that because of the protocell growth the volume is an increasing
function of time. Actually one can relate the volume to the amount of
container molecules via their density $V=C/\rho$ where $C$ denotes the integer
number of molecules forming the lipidic membrane. We will hereby
use the same symbol $X_i$ to denote both the $i$--th GMM and the integer
number of molecules of type $X_i$ in the system. 

For each 
reaction channel $R_{\mu}$ assume that there exists a scalar 
rate $c_{\mu}$ such that $c_{\mu}dt+o(dt)$ is the probability that
a random combination of molecules from channel $R_{\mu}$ will react in the
interval $[t,t+dt)$ within the volume $V(t)$.  

Let $h_{\mu}(Y)$ be the total number of possible distinct combinations of
molecules for a channel $R_{\mu}$ when the system is in state
$Y=(X_1,\dots,X_N,C)$, then we can define the {\em propensity}~\cite{LVTH2004}
of the reaction 
$R_{\mu}$ to be $a_{\mu}(Y)=h_{\mu}(Y)c_{\mu}$.

One can prove~\cite{Gillespie1997} that for a binary reaction the rate
$c_{\mu}$ can be written in 
the form $c_{\mu}=k_{\mu}/V$, where $k_{\mu}$ is a fixed constant. Similarly 
one can prove that for a reaction involving $n$ different species, we get:
$c_{\mu}=k_{\mu}/V^{n-1}$. And thus for a single molecule reaction, i.e. a
decay, we get $c_{\mu}=k_{\mu}$, namely independently from the volume.

Let us now assume that among the $m$ reactions, $Q_1$ involve one single
molecule, $Q_2$ are binary reactions, $Q_3$ are ternary reactions and so
on. Of course $Q_1+Q_2+\dots 
+Q_{N+1}=m$. We recall that we have $N$ GMMs and the container type molecule
$C$, hence $N+1$ species. For short we will denote $\mathcal{Q}_1$ the set of
indices $\mu$ for mono molecule reactions, and by $\mathcal{Q}$ the remaining
ones. Let us observe that in this way some coefficient
$a_{\mu}$,  will depend both on the system state $Y$ and  {on the}
time via the volume $V(t)$: $a_{\mu}(Y,t)$ for $\mu\in\mathcal{Q}$. 

 {More precisely} to study the time evolution of the system we need to determine the probability
$P_{\mu}(\tau|Y,t)d\tau$, that given 
the system in the state $Y=(X_1,\dots,X_n,C)$ at time $t$, then the next
reaction will occur in the infinitesimal time interval $(t+\tau,
t+\tau+d\tau)$ and it will be the reaction $R_{\mu}$. This probability will be
computed as
\begin{equation}
  \label{eq:Pdef}
 P_{\mu}(\tau|Y,t)d\tau  =  P_{not}(\tau|Y,t) \times
 a_{\mu}(Y,t+\tau)d\tau\, ,
\end{equation}
where $P_{not}(\tau|Y,t)$ is the probability that no reaction occurs in
$(t,t+\tau)$ given the state $Y$ at time $t$, whereas the rightmost term
denotes the probability to have a reaction $R_{\mu}$ in
$(t+\tau,t+\tau+d\tau)$ given the state $Y$ at time $t+\tau$.

To compute the first term $P_{not}$, let us take $s\in[t,t+\tau]$ and observe
that:
\begin{equation*}
  P_{not}(s+ds|Y,t)=P_{not}(s|Y,t)P_{not}(ds|Y,t+s)=P_{not}(s|Y,t)\left(1-\sum_{\mu}a_{\mu}(Y,t+s)ds\right)\, ,
\end{equation*}
being $1-\sum_{\mu}a_{\mu}(Y,t+s)ds$ the probability that no reaction will occur in $(t+s,t+s+ds)$ once we are in state
$Y$ at time $t+s$. Thus rewriting the previous difference equation as a
differential equation, passing 
to the limit $ds\rightarrow 0$, and observing that $P_{not}(0|Y,t)=1$, we get
the solution:
\begin{equation}
  \label{eq:Pnot}
  P_{not}(\tau|Y,t)={\rm exp}\left[ -A_{Q_1}(Y)\tau-\int_0^{\tau}A_Q(Y,s+t)\,
    ds\right]\, ,
\end{equation}
where
\begin{equation*}
  A_{Q_1}(Y)=\sum_{\mu\in\mathcal{Q}_1}a_{\mu}(Y) \quad \text{and}\quad
  A_{Q}(Y,s+t)=\sum_{\mu\in\mathcal{Q}}a_{\mu}(Y,s+t)\, .
\end{equation*}
The apparent asymmetry in the exponential term in~\eqref{eq:Pnot} is easily
recovered by observing that $A_{Q_1}(Y)\tau=\int_0^{\tau}A_{Q_1}(Y)ds$.

We can thus conclude that
\begin{equation}
  \label{eq:Pdeffin}
 P_{\mu}(\tau|Y,t)d\tau  = {\rm
   exp}\left[{-A_{Q_1}(Y)\tau-\int_t^{t+\tau}A_Q(Y,s)\, 
    ds}\right] a_{\mu}(Y,t+\tau)d\tau\, .
\end{equation}
Let us observe that the rightmost term is correctly $a_{\mu}(Y,t+\tau)$,
namely the system is still 
in the state $Y$ at time $t+\tau$, because no reaction has been produced in
$(t,t+\tau)$.

Let us recall that the volume enters in the previous relation via the function
$A_Q$, more explicitely one has
\begin{equation}
\label{eq:AQ}
  A_Q(Y,s)=\sum_{\mu\in
    \mathcal{Q}_2}\frac{h_{\mu}(Y)k_{\mu}}{V(s)}+\sum_{\mu\in
    \mathcal{Q}_3}\frac{h_{\mu}(Y)k_{\mu}}{(V(s))^2}+\dots+\sum_{\mu\in
    \mathcal{Q}_{N+1}}\frac{h_{\mu}(Y)k_{\mu}}{(V(s))^N}\, ,
\end{equation}
that can be rewritten in terms of $C$ molecules using the relation $C=\rho V$.
 So our method applies to a different problem with respect to the one 
considered in~\cite{LVTH2004}, in fact in our case the volume growth is not
imposed a priori but dynamically 
evolves accordingly to the reaction scheme,  {if $C$ is produced then $V$
  increases otherwise it will keep a constant value}, while in~\cite{LVTH2004} the
volume growth is an exogenous variable.

\section{The stochastic simulation algorithm in a growing volume}
\label{sec:algo}

Once we have the probability function $P_{\mu}(\tau|Y,t)$ we can build an
algorithm that reproduces the time evolution given by the model defined
above. 

Given 
the system in some state $Y$ at time $t$, we must determine the interval of
time $\tau$ and the reaction channel $R_{\mu}$ according to the probability
distribution function $P_{\mu}(\tau|Y,t)$, and finally update the state
$Y\rightarrow Y+\nu_{\mu}$, where $\nu_{\mu}$ is a stoichiometric vector
representing the increase and decrease of molecular abundance due to the
reaction $R_{\mu}$. This will be accomplished following the standard
approach by Gillespie~\cite{Gillespie1997} but taking care of the time
dependence of the propensities. We will thus need to compute the cumulative
probability distribution function and then make use of the inversion
method~\cite{GillespieBook}, to 
determine the channel $\mu$ and the next reaction time $\tau$, distributed
according to $P_{\mu}(\tau|Y,t)$.

From~\eqref{eq:Pdeffin} we can compute the {\em cumulative distribution
  function}
\begin{equation}
  \label{eq:cumpdf}
  F(\tau|Y,t)=\int_0^{\tau}\sum_{\mu}P_{\mu}(s|Y,t)\, ds\, ,
\end{equation}
providing the probability that any reaction will occur in $(t,t+\tau)$ starting
from the state $Y$ at time $t$. The function $F(\tau|Y,t)$ can be explicitely
computed by

\begin{proposition}
  \label{prop:cumpdf}
Under the above assumptions we have
\begin{equation}
  \label{eq:cumpdfexpl}
  F(\tau|Y,t)=1-{\rm exp}\left[{-A_{Q_1}(Y)\tau-\int_t^{t+\tau}A_Q(Y,s)\,
    ds}\right]\, .
\end{equation}
\end{proposition}

\proof
The first step is to use~\eqref{eq:Pdeffin} and perform a sum over all the
channels $\mu$ to rewrite~\eqref{eq:cumpdf} as
\begin{equation*}
  F(\tau|Y,t)=\int_0^{\tau}\left(A_{Q_1}(Y)+ A_Q(Y,t+s)\right)
    {\rm exp}\left[{-A_{Q_1}(Y)s-\int_t^{t+s}A_Q(Y,r)\, dr}\right]\, ds\, .
\end{equation*}
Then we can observe
that 
\begin{eqnarray*}
  &\,&\frac{\partial }{\partial s}\left({\rm
      exp}\left[{-A_{Q_1}(Y)s-\int_t^{t+s}A_Q(Y,r)\, 
    dr}\right]\right)=\\ &=&-\left(
    A_{Q_1}(Y)+A_Q(Y,t+s)\right){\rm
    exp}\left[{-A_{Q_1}(Y)s-\int_t^{t+s}A_Q(Y,r)\, dr}\right]\, , 
\end{eqnarray*}
and thus
\begin{eqnarray*}
  F(\tau|Y,t)&=&-\int_0^{\tau}\frac{\partial }{\partial
    s}\left({\rm exp}\left[{-A_{Q_1}(Y)s-\int_t^{t+s}A_Q(Y,r)\, 
    dr}\right] \right)\, ds\\
&=& 1-{\rm exp}\left[{-A_{Q_1}(Y)\tau-\int_t^{t+\tau}A_Q(Y,r)\, dr}\right] \, .
\end{eqnarray*}
\endproof

Once we have the cumulative distribution function we can obtain the value
$\tau$  by drawing a radom number $u_1$ from an uniform distribution in 
$[0,1]$ and then solve with respect to $\tau$ the implicit equation:
\begin{equation}
  \label{eq:fixtau}
  u_1=1-{\rm exp}\left[{-A_{Q_1}(Y)\tau-\int_t^{t+\tau}A_Q(Y,s)\, ds}\right]\, .
\end{equation}
Let us stress once again that this is not as straightforward as for the original
Gillespie~\cite{Gillespie1997} scheme, or the simplified one presented
in~\cite{LVTH2004}, because of the time dependence of $A_Q$ via the
volume. One can 
nevertheless found suitable approximation for the integral, this will be the
goal of the next sections.

\subsection{The adiabatic assumption}
\label{ssec:adiab}

Let assume that $\tau$ is very small, or which is equivalent, that the time
scale of 
the chemical reactions involving the GMMs is much faster than the production of
container molecules, hence the volume growth is very slow compared with the
production of the chemicals $X_i$.

Under this hypothesis one can assume that in the interval $(t,t+\tau)$ the
volume doesn't vary and thus one can made the following approximation
\begin{equation}
  \label{eq:adiaint}
  \int_t^{t+\tau}A_Q(Y,s)\, ds\sim A_Q(Y,t)\tau\, .
\end{equation}

One can thus explicitely solve equation~\eqref{eq:fixtau} to get:
\begin{equation}
  \label{eq:tauadia}
  \tau_{Gill}=-\frac{1}{A_{Q_1}(Y)+A_Q(Y,t)}\log (1-u_1)\, ,
\end{equation}
that is the standard Gillespie result except now that $A_Q(Y,t)$ depends on
time and as long the volume increases, then the contribution arising form $A_Q(Y,t)$ mights become smaller
because $A_Q\sim 1/V$.

\subsection{The next order correction}
\label{ssec:noc}

One can obtain a somehow better estimate valid in the case of comparable time
scales for the reactions involving GMM and the container growth. The idea is to
compute the integral in Eq.~\eqref{eq:fixtau} using the following approximation:
\begin{eqnarray}
  \label{eq:noc}
  \int_t^{t+\tau}A_Q(Y,s)\, ds &=& \int_0^{\tau}A_Q(Y,t+s)\, ds  =
  \int_0^{\tau}\left(A_Q(Y,t)+\frac{\partial {A}_Q(Y,t)}{\partial t}s+\dots
  \right)\, ds \notag\\ 
&=& A_Q(Y,t)\tau +\frac{\partial {A}_Q(Y,t)}{\partial
  t}\frac{\tau^2}{2}+\mathcal{O}(\tau^3)\, .
\end{eqnarray}
where ${\partial {A}_Q(Y,t)}/{\partial t}$ can be obtained using the
definition~\eqref{eq:AQ} and expressing the volume in terms of $C=V(t)\rho$,
namely:  
\begin{equation*}
  \frac{\partial {A}_Q(Y,t)}{\partial t}=-\frac{\dot{C}}{C}\left(\sum_{\mu\in
    \mathcal{Q}_2}\frac{h_{\mu}(Y)k_{\mu}}{C(t)}+2\sum_{\mu\in
    \mathcal{Q}_3}\frac{h_{\mu}(Y)k_{\mu}}{(C(t))^2}+\dots+N\sum_{\mu\in
    \mathcal{Q}_{N+1}}\frac{h_{\mu}(Y)k_{\mu}}{(C(t))^N}\right)\, .  
\end{equation*}
To compute ${\dot C/ C}$ we make the assumption that in a very short time
interval, as the one we are interested in, the deterministic growth of the
container is a good approximation for the stochastic underlying mechanism;
this implies that we can use~\eqref{eq:ode}
\begin{equation*}
  \frac{\dot{C}}{C}=\left(\frac{C(t)}{\rho}\right)^{\beta-1}\frac{\vec{\alpha}\cdot
  \vec{X}(t)}{C(t)}\, .
\end{equation*}

Inserting the previous result into~\eqref{eq:noc} and finally
solving~\eqref{eq:fixtau} with respect to $\tau$, we can compute the next
reaction time up to correction of the order 
of $\tau^3$, as:
\begin{equation}
  \label{eq:taunoc}
\tau_{Gill} = \frac{-(A_{Q_1}(Y)+A_{Q}(Y,t))+\sqrt{(A_{Q_1}(Y)+A_{Q}(Y,t))^2-2\log
    (1-u_1) \dot{A}_Q(Y,t)}}{\dot{A}_Q(Y,t)}\, ,
\end{equation}
where we wrote for short $\dot{A}_Q(Y,t)=\partial A_Q(Y,t)/\partial t$ and we
selected the positive square root in such a way in the limit 
$\dot{A}_Q(Y,t)\rightarrow 0$ we recover the previous
solution~\eqref{eq:tauadia}. 

\begin{remark}[On the existence of $\tau_{Gill}$]
 In the case of variable volume a new phenomenon can arise: the volume growth
 can be so fast that no reaction can occur in the interval $(t,t+\tau+d\tau)$
 for 
 any $\tau$. Mathematically this translates into
 a sign condition for the term under square root in~\eqref{eq:taunoc}, if:
\begin{equation}
\label{eq:existtgill}
\log(1-u_1) < (A_{Q_1}(Y)+A_{Q}(Y,t))^2/(2\dot{A}_Q(Y,t))\, ,
\end{equation}
then equation~\eqref{eq:fixtau} has no real solution.

This can be geometrically interpreted as follows. The
relation~\eqref{eq:fixtau} determines $\tau_{Gill}$ as the intersection of the
parabola $-A_{Q_1}(Y)-A_{Q}(Y,t)\tau-\dot{A}_Q(Y,t)\tau^2/2$ with the
horizontal line $\log(1-u_1)$, which is negative because $u_1\in(0,1)$. Such
parabola intersect the $y$-axis at $\tau_1=0$ and
$\tau_2=-2(A_{Q_1}(Y)+A_{Q}(Y,t))/\dot{A}_Q(Y,t)>0$ and it is concave. Then
its absolute (negative) minimum is reached at the vertex $\tau_V=(t_1+t_2)/2$
and its value is $(A_{Q_1}(Y)+A_{Q}(Y,t))^2/(2\dot{A}_Q(Y,t))$ and it is
negative because $\dot{A}_Q(Y,t)$ is negative. Hence if the
horizontal line is below this value,
i.e. condition~\eqref{eq:existtgill} is verified, the parabola and the line do
not have any 
real intersections (see Fig.~\ref{fig:nointer}).
 \begin{center}
 \begin{figure}[ht]
  \begin{center}
  \includegraphics[scale=0.25]{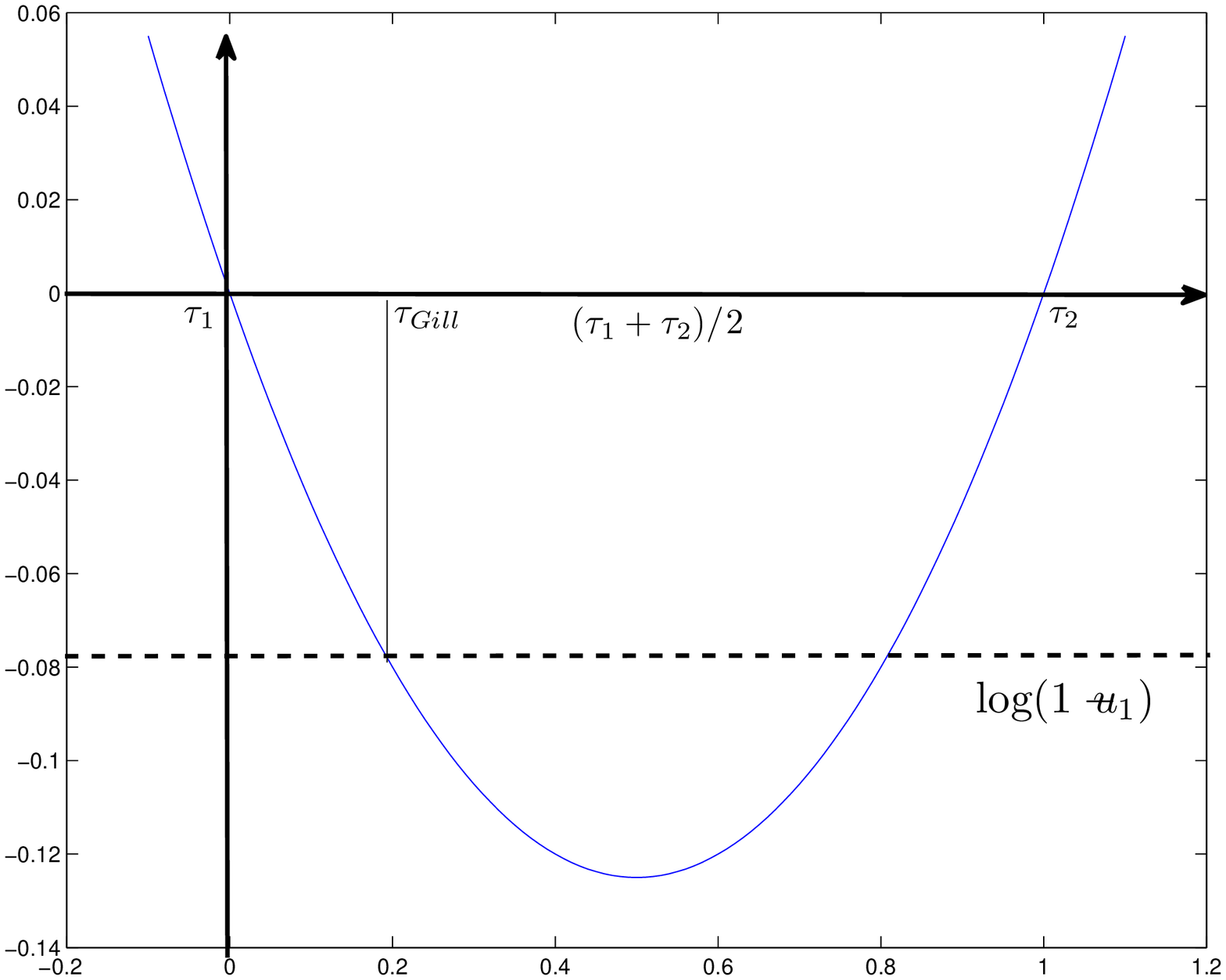}\quad \includegraphics[scale=0.25]{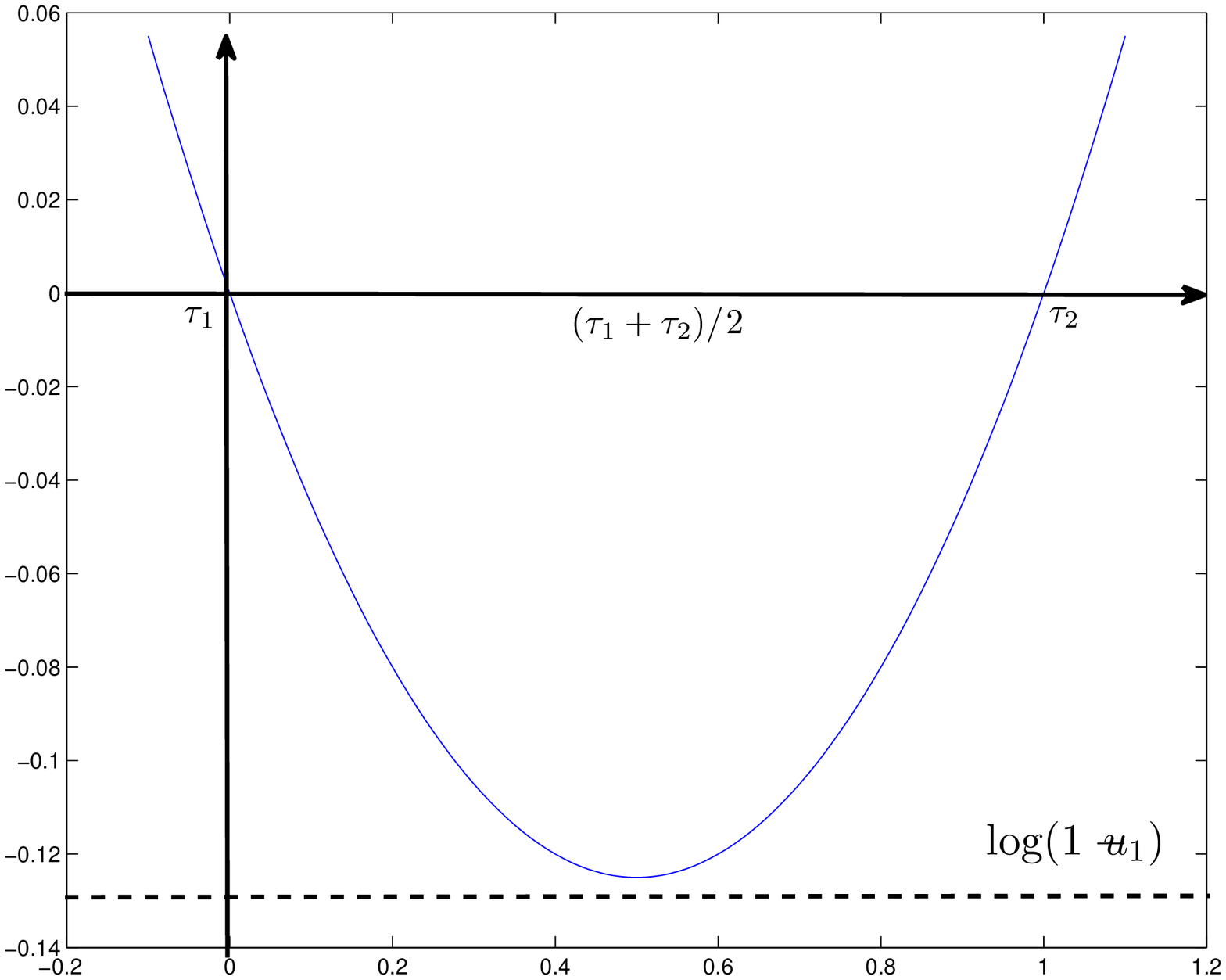} 
  \caption{Geometrical interpretation of the existence of the next reaction
    time $\tau_{Gill}$. Left panel : $\tau_{Gill}$ is the smallest
    intersection between the parabola and the horizontal line
    $\log(1-u_1)$. Right panel $\tau_{Gill}$ doesn't exist, the horizontal
    line is located below the minimum of the parabola.} 
  \label{fig:nointer}
  \end{center}
 \end{figure}
 \end{center}

Let us also observe that, whenever it exists, $\tau_{Gill}$ is always
positive as it should be. In the case of a protocell the non existence of such next reaction time could be translated into the death by dilution of the protocell.
\end{remark}

\subsection{The next reaction channel}
\label{ssec:rchan}

 {Whenever} the next reaction time does exist, the next reaction channel is
determined using the classical Gillespie method, namely by drawing a second
uniformly distributed random number $u_2\in[0,1]$ and fix the channel $\mu$
such that: 
\begin{equation}
  \label{eq:musel}
  \sum_{{\nu}=1}^{\mu-1} a_{\nu}(Y,t+\tau)\leq u_2 a_0(Y,t+\tau)\leq \sum_{{\nu}=1}^{\mu} a_{\nu}(Y,t+\tau)\, ,
\end{equation}
where $a_0(Y,t+\tau) =A_{Q_1}(Y)+A_{Q}(Y,t+\tau)=\sum_{\nu=1}^m a_{\nu}(Y,t+\tau)$.

\begin{remark}
\label{rem:novoldep}
  Let us observe that if all the reactions involve the same number of
  chemicals, then the determination of which reaction channel $\mu$  {will
    be activated} in the next reaction, 
  doesn't depend on the volume which factorizes out from~\eqref{eq:musel}. In
  fact assuming all the reactions to involve $p$ chemical, we obtain by
  definition
  \begin{equation*}
    a_{\nu}(Y,t+\tau)=\frac{h_{\nu}(Y)k_{\nu}}{\left[V(t+\tau)\right]^p}\quad \forall
    \nu\in\{1,\dots,m\}\, ,
  \end{equation*}
and thus~\eqref{eq:musel} rewrites:
\begin{equation*}
  \sum_{{\nu}=1}^{\mu-1} \frac{h_{\nu}(Y)k_{\nu}}{\left[V(t+\tau)\right]^p}\leq u_2
  \sum_{\nu=1}^m \frac{h_{\nu}(Y)k_{\nu}}{\left[V(t+\tau)\right]^p}\leq
  \sum_{{\nu}=1}^{\mu} \frac{h_{\nu}(Y)k_{\nu}}{\left[V(t+\tau)\right]^p}\, , 
\end{equation*}
which is clearly independent of the volume value $V$.
\end{remark}

\section{Some applications}
\label{sec:app}

The aim of this section is to provide some applications of the previous
algorithm to the study of the evolution of a protocell.

\subsection{One single Genetic Memory Molecule}
\label{ssec:1ggm}

The simplest model is the one where only one GMM specie is present in the
protocell~\cite{SerraAlife} and thus only two chemical channels are
active:
\begin{eqnarray}
  \label{eq:1ggmschema}
 \text{channel $1$, $R_1$}& :\qquad &X+P_1 \autorightarrow{$\eta$}{} 2X \notag
 \\ 
 \text{channel $2$, $R_2$}& :\qquad   &X+L_1 \autorightarrow{$\alpha$}{} X+C
 \, , 
\end{eqnarray}
where $P_1$ and $L_1$ are, respectively, precursors of GMM, i.e. nucleotide,
and precursors of amphiphiles.

One can thus compute the propensities in the state $Y=(X,C)$ at time
$t$: 
\begin{equation}
  \label{eq:prop1gmm}
  a_1(X,C,t)=h_1(X,C)\frac{\eta}{V(t)}=\eta\frac{P_1 X}{V(t)} \quad
  \text{and} \quad 
  a_2(X,C,t)=h_2(X,C)\frac{\alpha}{V(t)}=\alpha\frac{L_1 X}{V(t)}\, ;
\end{equation}
let us observe that we assume that precursors are buffered and thus they are
constant. 

Because system~\eqref{eq:1ggmschema} contains only bimolecular reactions, all
the propensities are time dependent, hence $A_{Q_1}=0$ and
$A_{Q}=a_1(X,C,t)+a_2(X,C,t)=(P_1\eta+L_1\alpha)X/V(t)$,
thus~\eqref{eq:fixtau} simplifies into
\begin{equation*}
  u_1=1-{\rm exp}\left[{-\int_t^{t+\tau}A_Q(Y,s)\, ds}\right]\, ,
\end{equation*}
whose second order solution~\eqref{eq:taunoc} is given by
\begin{equation*}
\tau_{Gill} = \frac{-A_{Q}(Y,t)+\sqrt{(A_{Q}(Y,t))^2-2\log
    (1-u_1) \dot{A}_Q(Y,t)}}{\dot{A}_Q(Y,t)}\, ,
\end{equation*}
and
\begin{eqnarray*}
  \frac{\partial {A}_Q(X,C,t)}{\partial
    t}&=&-\frac{\dot{V}(t)}{V(t)}\left(\frac{P_1\eta X}{V(t)}+\frac{L_1\alpha X}{V(t)}\right)\Big\rvert_{V(t)=C(t)/\rho}=-
\left(\frac{C}{\rho}\right)^{\beta-1}\frac{\rho L_1\alpha X^2}{C^2}\left(P_1
  \eta+L_1 \alpha\right)\, .      
\end{eqnarray*}
So we can finally obtain
\begin{equation*}
\tau_{Gill} = \frac{C}{L_1 \alpha
  X}\left(\frac{\rho}{C}\right)^{\beta-1}-\sqrt{\left[\frac{C}{L_1 \alpha
  X}\left(\frac{\rho}{C}\right)^{\beta-1}\right]^2+2\frac{C^2}{L_1 \alpha \rho X^2(P_1\eta+L_1\alpha)}\log(1-u_1)}\, ,
\end{equation*}
provided
\begin{equation*}
  \log(1-u_1) \geq -\frac{\rho}{2\alpha}
  \left(\frac{\rho}{c}\right)^{2(\beta-1)} (P_1\eta +L_1\alpha)\, .
\end{equation*}

Which reaction channel $\mu$ will active in the time interval $[t,t+\tau]$ can
be obtained according to :  
\begin{eqnarray*}
   \text{ if } u_2\frac{(P_1 \eta+L_1 \alpha)X}{V}\leq
  \frac{P_1 \eta X}{V} &\text{ namely } 0\leq u_2\leq
  \frac{P_1 \eta}{P_1 \eta+L_1 \alpha} &\text{then }\mu=1 \\
 \text{ if } \frac{P_1 \eta X}{V} < u_2 \frac{(P_1 \eta + L_1 \alpha)X}{V}\leq
 \frac{(P_1 \eta+L_1 \alpha)X}{V} &\text{ namely }
 \frac{P_1 \eta}{P_1 \eta+L_1 \alpha}< u_2 \leq 1 &\text{then }\mu=2  \, .
\end{eqnarray*}

Let us observe that according to remark~\ref{rem:novoldep}, the choice of
$\mu$ doesn't depend on the volume, because only binary reactions are present.

Let $C_0$ be the initial amount of container molecules, then we assume that
once $C(\bar{t})=2C_0$ the protocell splits into two offspring, almost halving
the 
GMM amount. More precisely we assume that the first offspring will get a number
of GMMs drawn according to a Binomial distribution with parameter $p=1/2$ and
$n=X(\bar{t})$. From this step, for technical reason, only one randomly chosen
offspring will be studied  {during each generation.}

\begin{center}
 \begin{figure}[ht]
  \begin{center}
	  \includegraphics[scale=0.25]{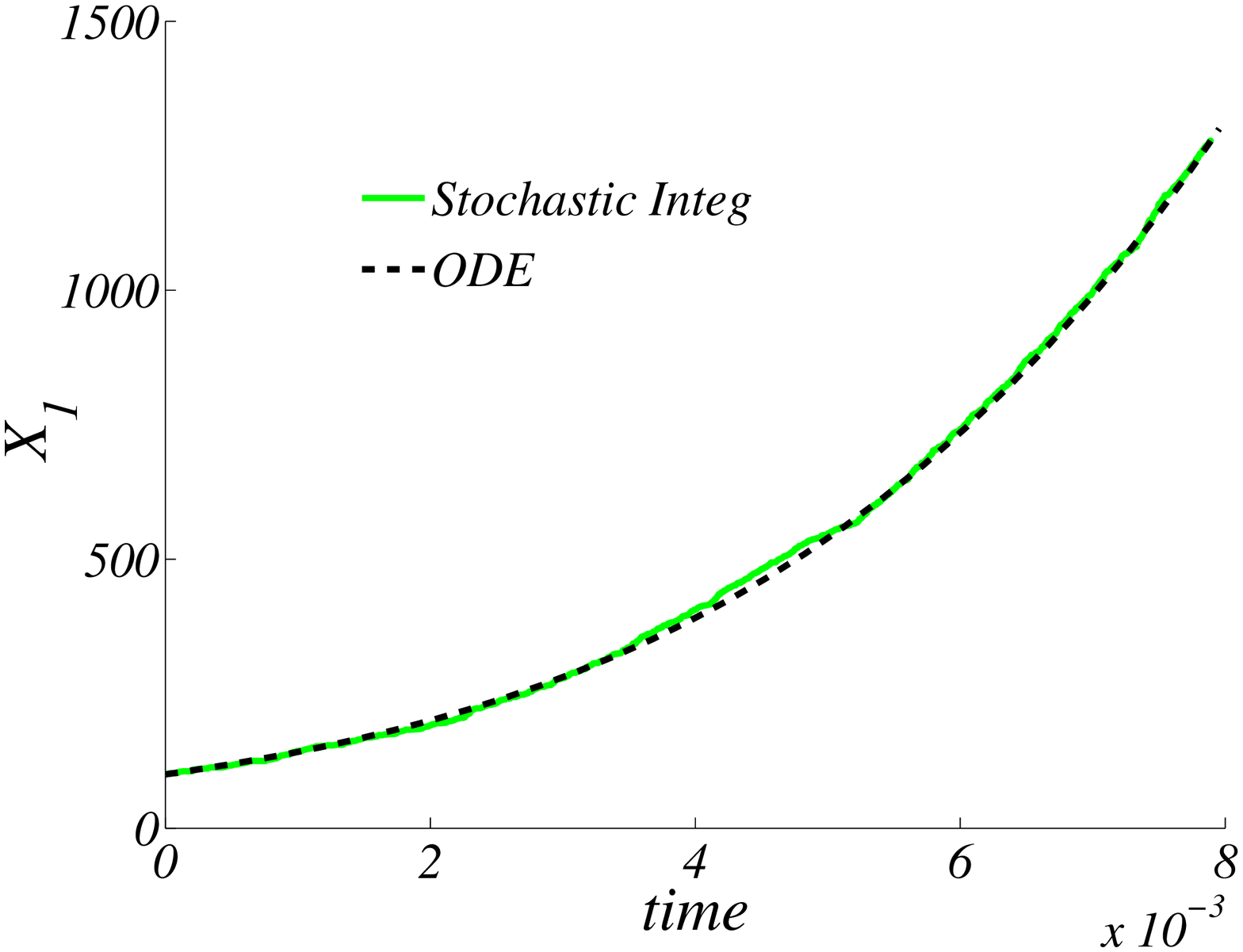}\quad \includegraphics[scale=0.25]{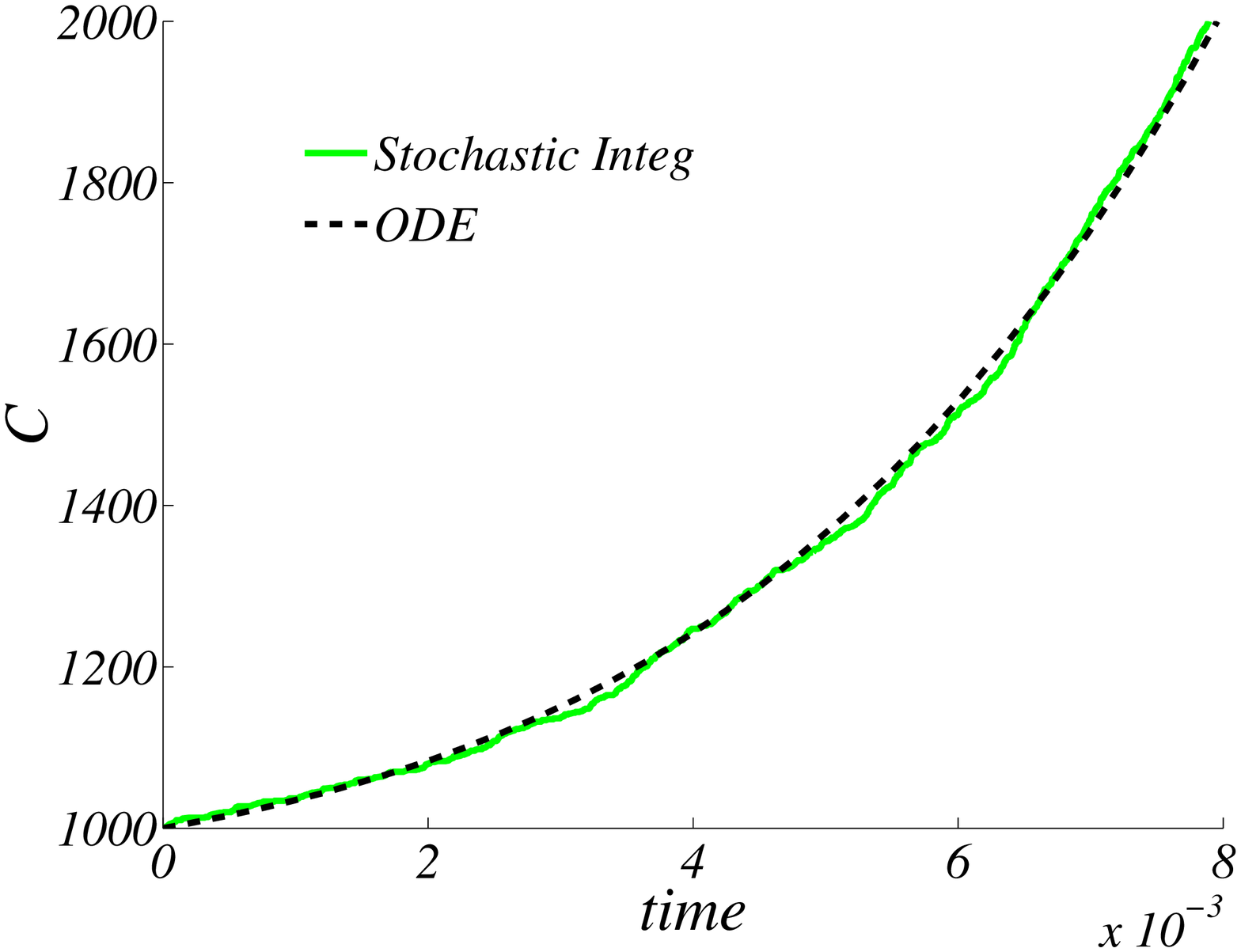}b 
  \caption{Stochastic vs ODE SRM protocell~\eqref{eq:ode}. Case of one GMM, left panel the
    time evolution of the amount of GMM, right panel the time evolution of the
    amount of
    $C$. Parameters are : 
    $\eta=1$, $\alpha=1$, $L1=500$, $P1=600$, $X_1(0)=100$, $C(0)=1000$,
    $\rho=200$ and $\beta=2/3$.} 
  \label{fig:XCSRM1divlarge}
  \end{center}
 \end{figure}
 \end{center}

In Fig.~\ref{fig:XCSRM1divlarge} we report a comparison between the
deterministic~\eqref{eq:ode} and the stochastic dynamics, under the adiabatic assumption for
$\tau_{Gill}$, corresponding to the continuous
growth phase of the container between two successive divisions. As one should
expect, a system composed by a large number of molecules exhibits small
stochastic fluctuations whose average is not too far from the dynamics
described by the deterministic model.

\begin{center}
 \begin{figure}[ht]
  \begin{center}
  \includegraphics[scale=0.23]{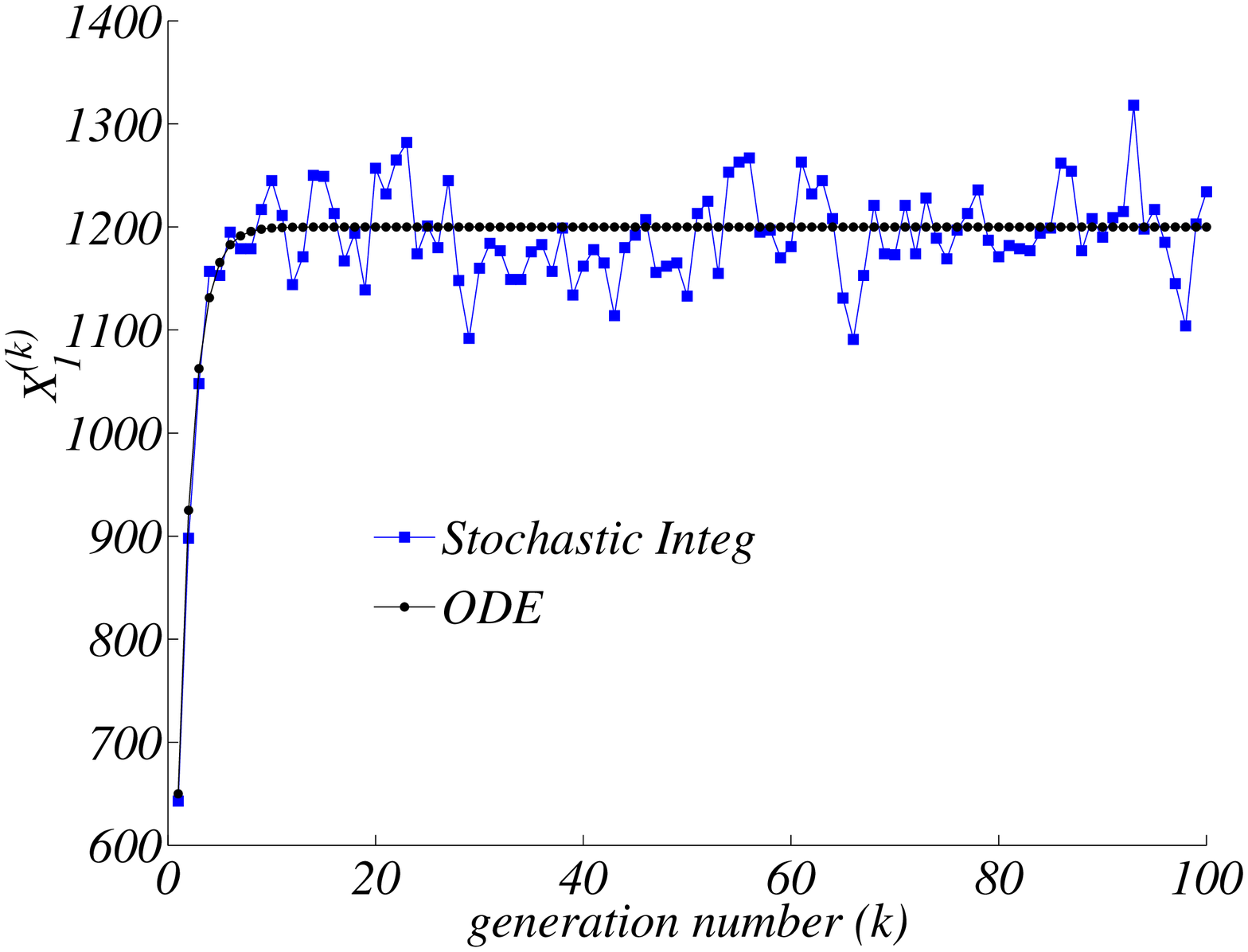}(a)\quad 
  \includegraphics[scale=0.23]{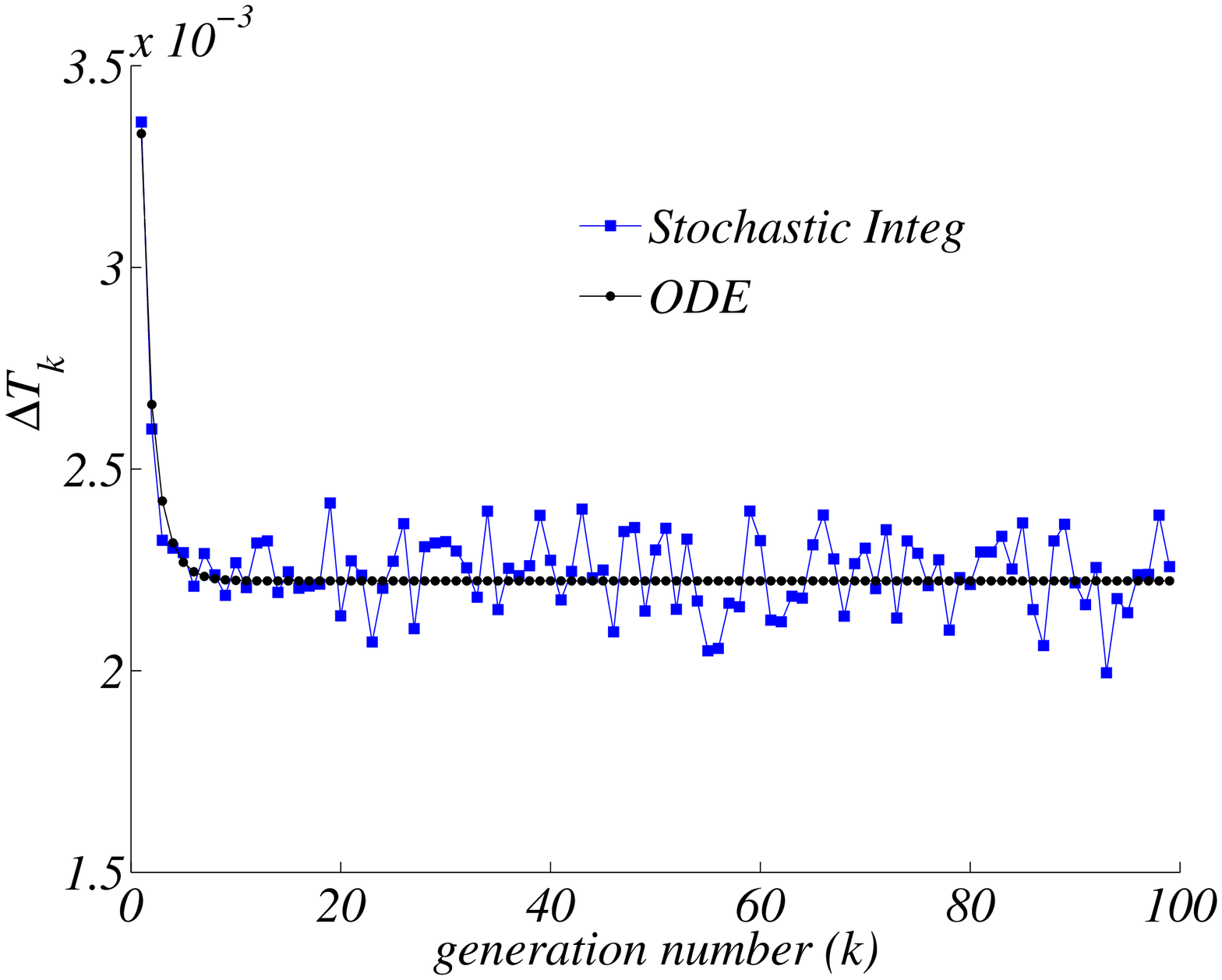}(b) 
  \caption{Stochastic vs ODE SRM protocell~\eqref{eq:ode}. Case of one GMM, left panel the
    amount of GMM at the beginning of each division cycle, right panel the
    division time as a function of the number of elapsed divisions. Parameters
    are : $\eta=1$, $\alpha=1$, $L1=500$, $P1=600$, $X_1(0)=100$, $C(0)=1000$, 
    $\rho=200$ and $\beta=2/3$.} 
  \label{fig:XCSRMNdivlarge}
  \end{center}
 \end{figure}
 \end{center}

In Fig.~\ref{fig:XCSRMNdivlarge} we report the amount of GMM, $X^{(k)}$ (panel
a), at
the beginning of each protocell cycle and the duplication time (panel b),
namely the 
interval of time needed to double the amount of 
$C$ molecules, for both the stochastic and
deterministic models. Once again one can clearly
observe the small fluctuations of the stochastic system around the value obtained by the numerical integration of the deterministic 
description, Eq.~\eqref{eq:ode}.  {Let us observe that these fluctuations
  are due to the stochastic integrator scheme and also on the division
  mechanism.}

\begin{center}
 \begin{figure}[ht]
  \begin{center}
  \includegraphics[scale=0.25]{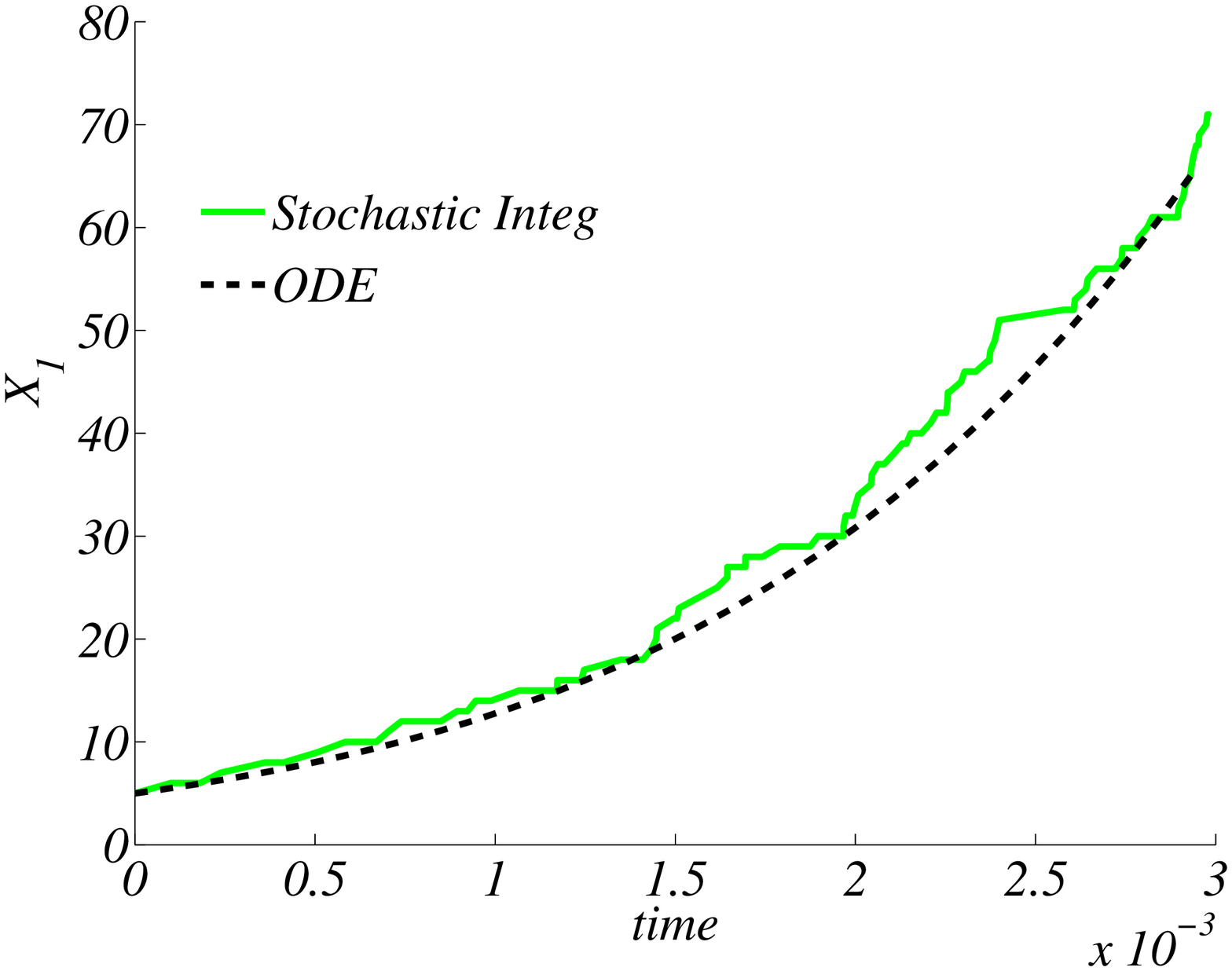}\quad \includegraphics[scale=0.25]{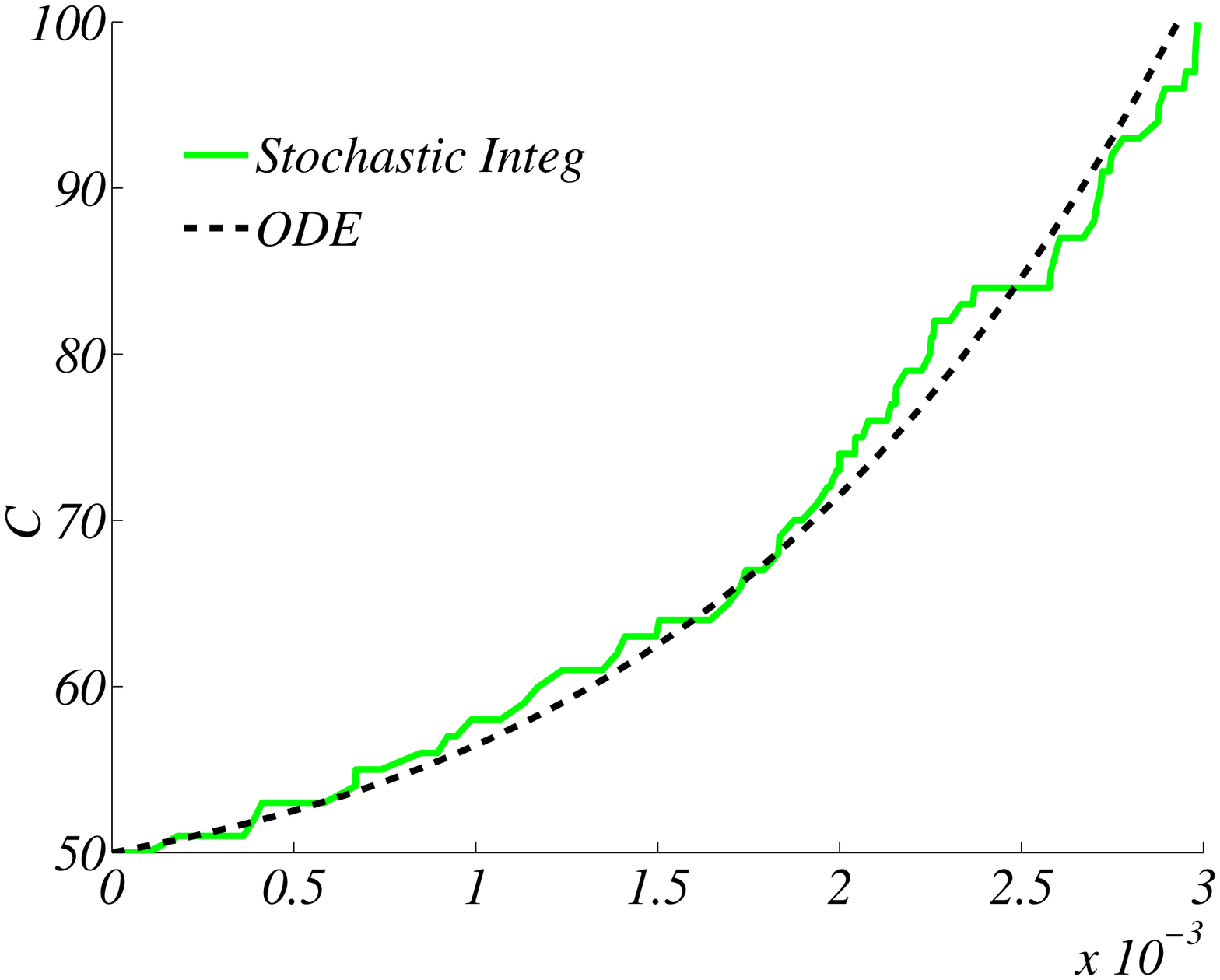} 
  \caption{Stochastic vs ODE SRM protocell~\eqref{eq:ode}. Case of one GMM, left panel the
    time evolution of the amount of GMM, right panel the time evolution of the
    amount of
    $C$. Parameters are : 
    $\eta=1$, $\alpha=1$, $L1=500$, $P1=600$, $X_1(0)=5$, $C(0)=50$,
    $\rho=200$ and $\beta=2/3$.} 
  \label{fig:XCSRM1divsmall}
  \end{center}
 \end{figure}
 \end{center}

\begin{center}
 \begin{figure}[ht]
  \begin{center}
  \includegraphics[scale=0.25]{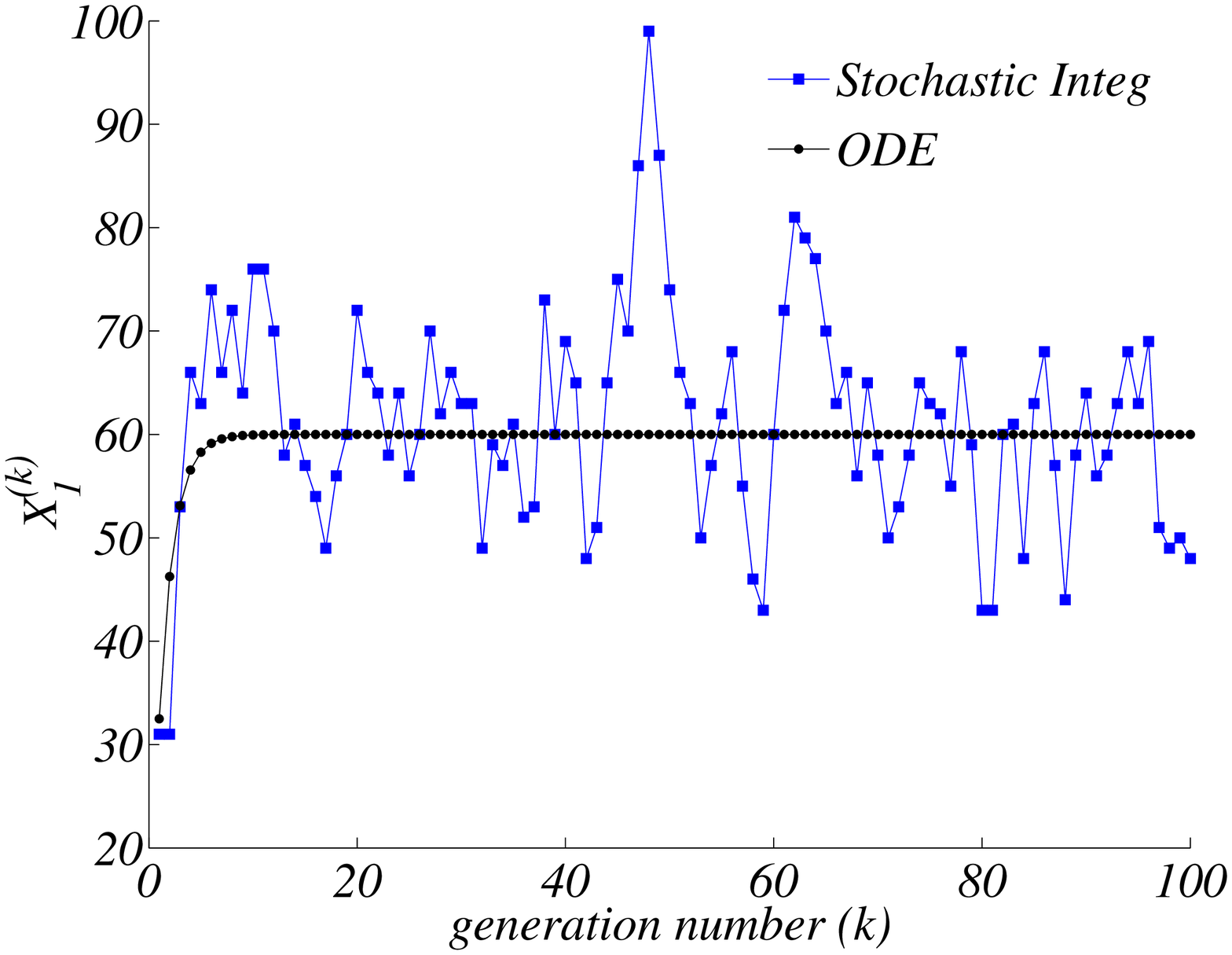}\quad \includegraphics[scale=0.25]{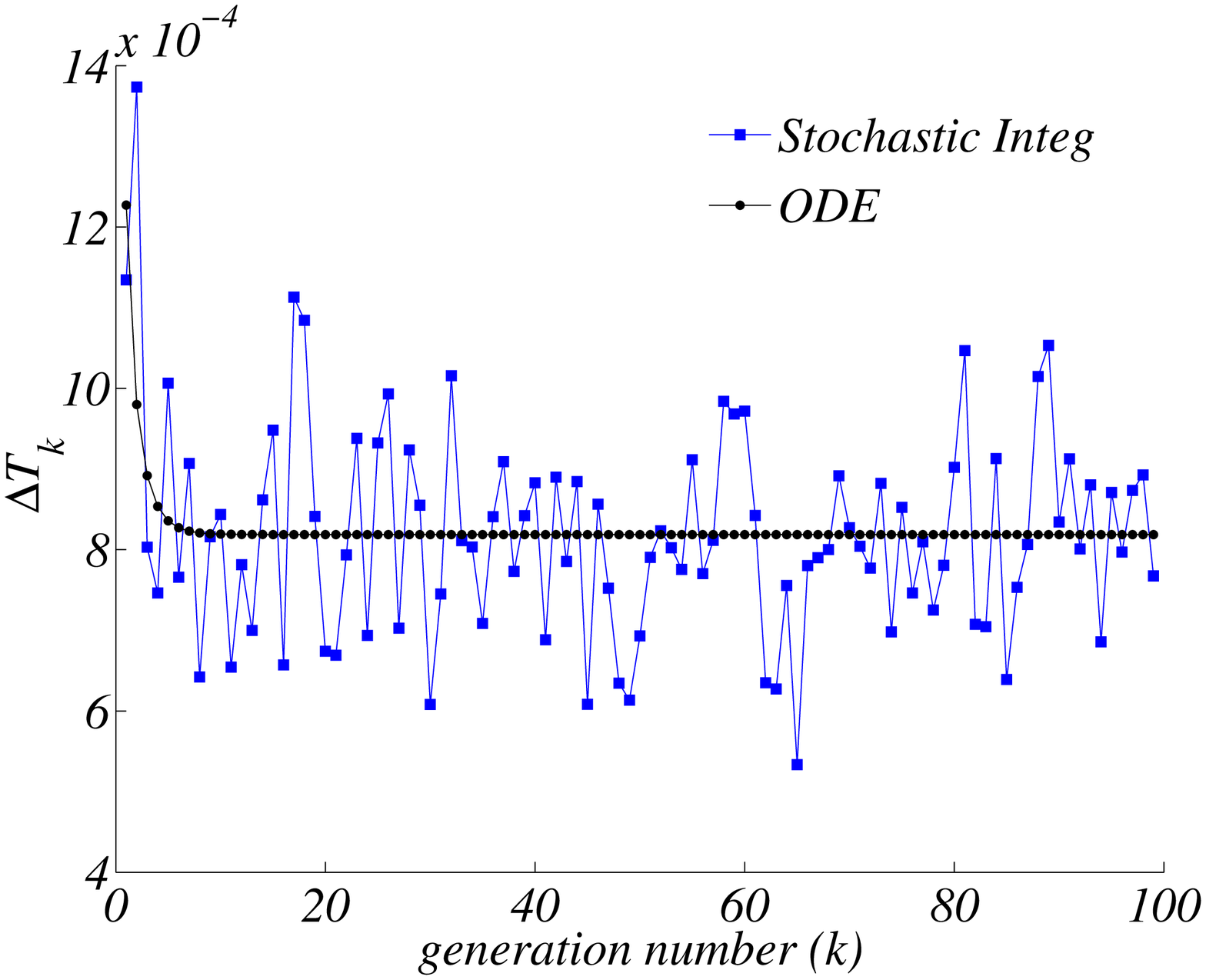} 
  \caption{Stochastic vs ODE SRM protocell~\eqref{eq:ode}. Case of one GMM, left panel the
    amount of GMM at the beginning of each division cycle, right panel the
    division time as a function of the number of elapsed divisions. Parameters
    are : $\eta=1$, $\alpha=1$, $L1=500$, $P1=600$, $X_1(0)=5$, $C(0)=50$, 
    $\rho=200$ and $\beta=2/3$.} 
  \label{fig:XCSRMNdivsmall}
  \end{center}
 \end{figure}
 \end{center}

We are now interested in studying the fluctuations dependence on the amount of
 molecules. We already know that for a sufficiently large number of
molecules the stochastic dynamics follows closely the deterministic one and
thus the fluctuations are small. On the other hand, one should expect that
when the number of molecules decreases, then the fluctuation will rise and the
system behavior could not be completely described by means of a deterministic
approach. This is confirmed by Fig.~\ref{fig:XCSRM1divsmall} and
Fig.~\ref{fig:XCSRMNdivsmall}, where we can observe that a model composed by a
small number of initial molecules, $20$ times lesser than in the model
presented in Fig.~\ref{fig:XCSRM1divlarge} exhibits larger stochastic
fluctuations.

\begin{center}
 \begin{figure}[ht]
  \begin{center}
  \includegraphics[scale=0.27]{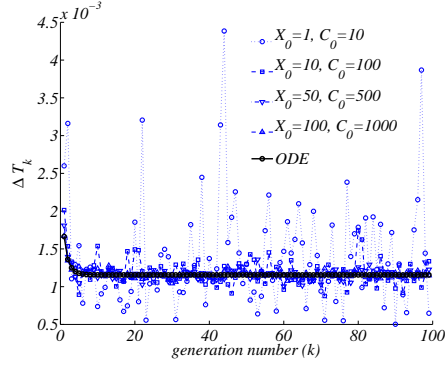}
  \caption{Fluctuation dependence on the initial conditions. We report the
    division times as a function of the number of elapsed divisions, for $5$
    different protocells models. 
Protocell $\circ$ : $X_1(0)=5$, $C(0)=10$, protocell $\square$: $X_1(0)=10$,
    $C(0)=100$, protocell $\triangledown$: $X_1(0)=50$, $C(0)=500$,
    protocell $\triangle$: $X_1(0)=100$, $C(0)=1000$. The black line
    denotes the deterministic protocell. All the remaining parameters have
    been fixed to: $\eta=1$, $\alpha=1$, $L1=500$, $P1=600$, $\rho=100$ and
    $\beta=1$.}  
  \label{fig:XCSRMscreening}
  \end{center}
 \end{figure}
 \end{center}

In Fig.~\ref{fig:XCSRMscreening} we summarize the results of several protocell
models each one with a different amount of initial molecules, in order to
appreciate the influence of the latter on the stochastic
fluctuations. To compare with, we also report the case of the deterministic
model. Because the kinetic constants are kept constant, the analytical theory
for the deterministic model ensures that the division time doesn't
vary~\cite{CarlettiJTB}. Nevertheless the fewer 
is the initial amount of $X_{0}$ and $C_{0}$, the larger are the fluctuations
present in the stochastic integration. 

\begin{center}
 \begin{figure}[ht]
  \begin{center}
  \includegraphics[scale=0.30]{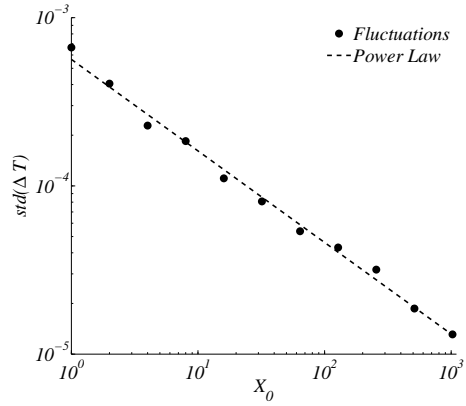}
  \caption{Fluctuation dependence on the initial conditions. We report the
standard deviation of the protocell division time as a function of the initial
amount of molecules $X_{0}$ ($\bullet$) and a linear best fit, whose slope
is $=-0.54\pm 0.03$. Parameters are:
$X(0)=2^{n}$ with $n=0, ..., 10$,~$C(0)=10X(0)$, $\eta=1$, $\alpha=1$,
$L1=500$, $P1=600$, $\rho=100$ and $\beta=1$.}  
  \label{fig:XCSRMSTDDeltaT}
  \end{center}
 \end{figure}
 \end{center}

To get a more complete understanding of the fluctuations dependence, we decided
to measure them using the standard deviation of the protocell division time
 {(after a sufficiently long transient phase)}. 
In Fig.~\ref{fig:XCSRMSTDDeltaT} we report the standard deviation of the
division time $\Delta T$ as a function of the initial amount of molecules.
As expected the fluctuations strength decreases rapidly as soon as the number
of molecules increases and the relation can be very well approximated by a power law distribution with exponent $-0.54\pm 0.03$ (linear best fit).

\subsection{Two non--interacting Genetic Memory Molecules}
\label{ssec:2nnintggm}

A slightly more sophisticated model can be obtained by considering two linear
non interacting GMMs. The system can be described by the following chemical reactions:
 
\begin{eqnarray}
  \label{eq:1ggmschema2}
 \text{channel $1$, $R_1$}& :\qquad &X_1+P_1 \autorightarrow{$\eta_1$}{} 2X_1 \notag \\ 
 \text{channel $2$, $R_2$}& :\qquad &X_1+L_1 \autorightarrow{$\alpha_1$}{} X_1+C \notag \\ 
 \text{channel $3$, $R_3$}& :\qquad &X_2+P_2 \autorightarrow{$\eta_2$}{} 2X_2 \notag \\ 
 \text{channel $4$, $R_4$}& :\qquad &X_2+L_2 \autorightarrow{$\alpha_2$}{} X_2+C
 \notag \, ,
\end{eqnarray}
where $P_i$ and $L_i$ are, respectively, precursors of the $i$--th GMM, i.e. nucleotide,
and precursors of amphiphiles used by the $i$--th GMM to build a $C$ molecule.

\begin{center}
 \begin{figure}[ht]
  \begin{center}
  \includegraphics[scale=0.2]{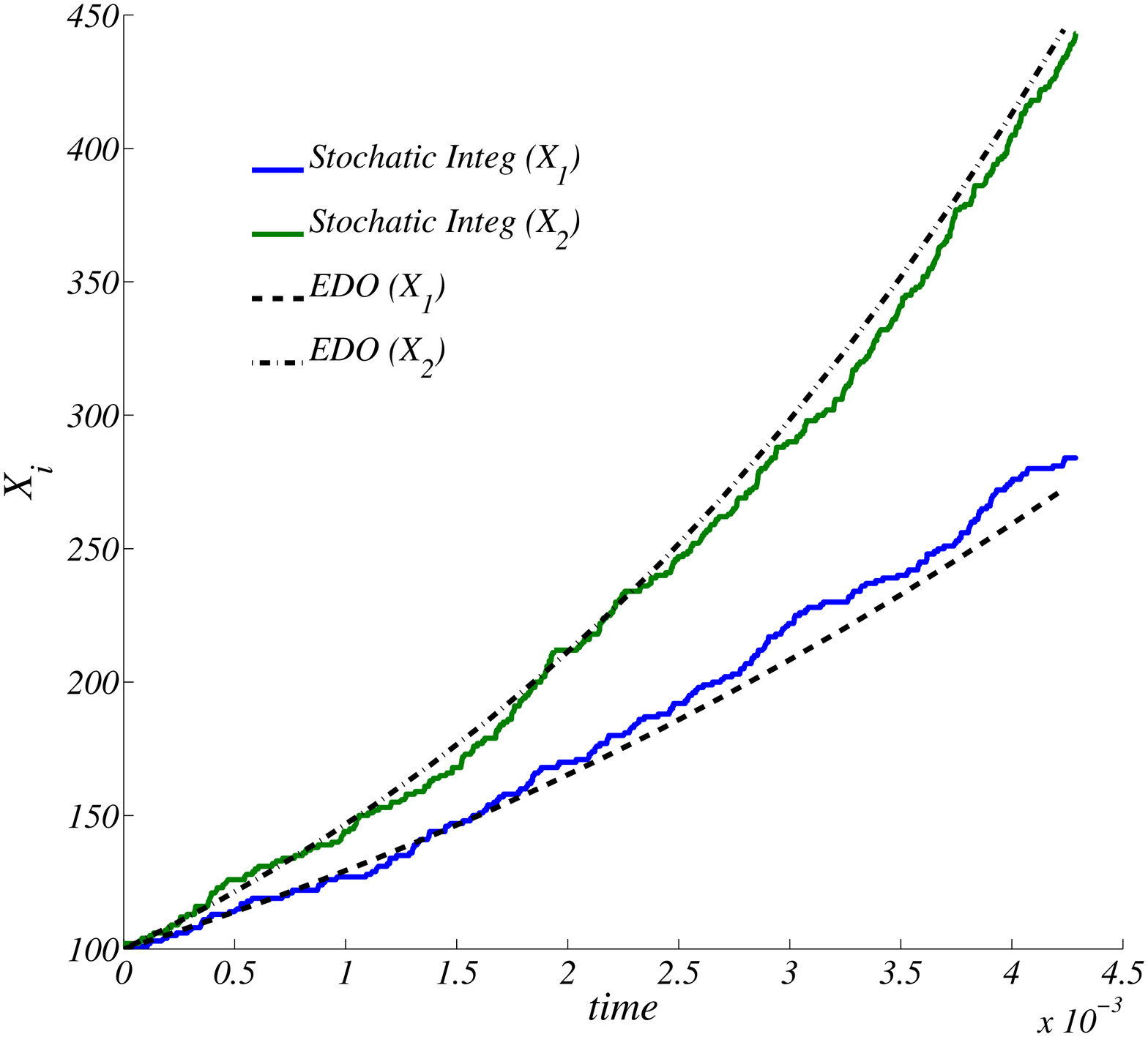}\quad \includegraphics[scale=0.2]{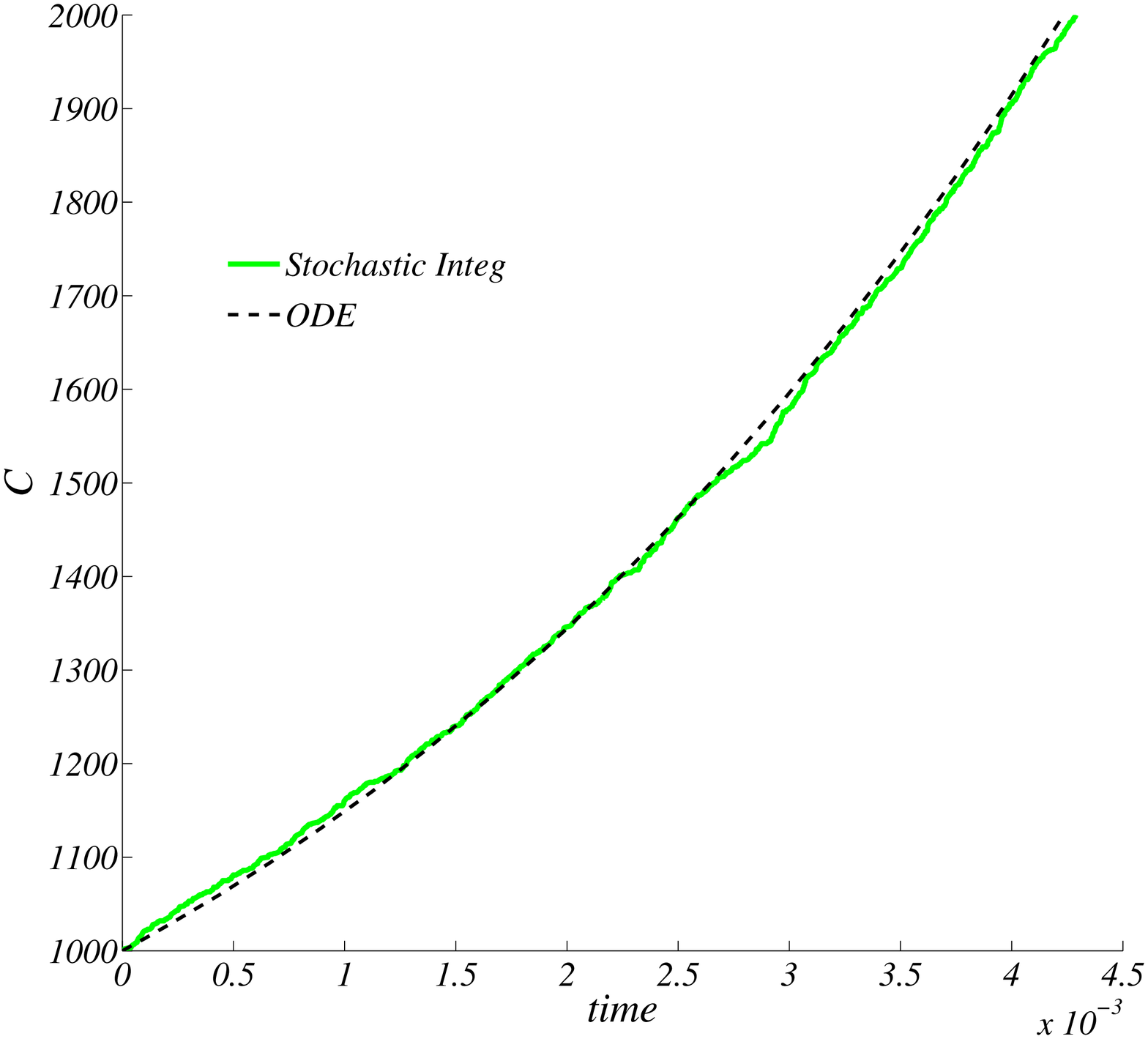} 
  \caption{Stochastic vs ODE SRM protocell~\eqref{eq:ode}. Case of two GMMs, left panel the
    time evolution of the amount of GMM during a division cycle, right panel
    the time evolution of the amount of $C$ molecules. Parameters
    are : $\eta_1=\eta_2=1$, $\alpha_1=\alpha_2=2$, $L1=500$, $L2=600$,
    $P1=600$, $P_2=670$, $X_1(0)=X_2(0)=100$, $C(0)=1000$, $\rho=200$ and $\beta=2/3$.} 
  \label{fig:XYCSRM1divlarge}
  \end{center}
 \end{figure}
 \end{center}

\begin{center}
 \begin{figure}[ht]
  \begin{center}
  \includegraphics[scale=0.2]{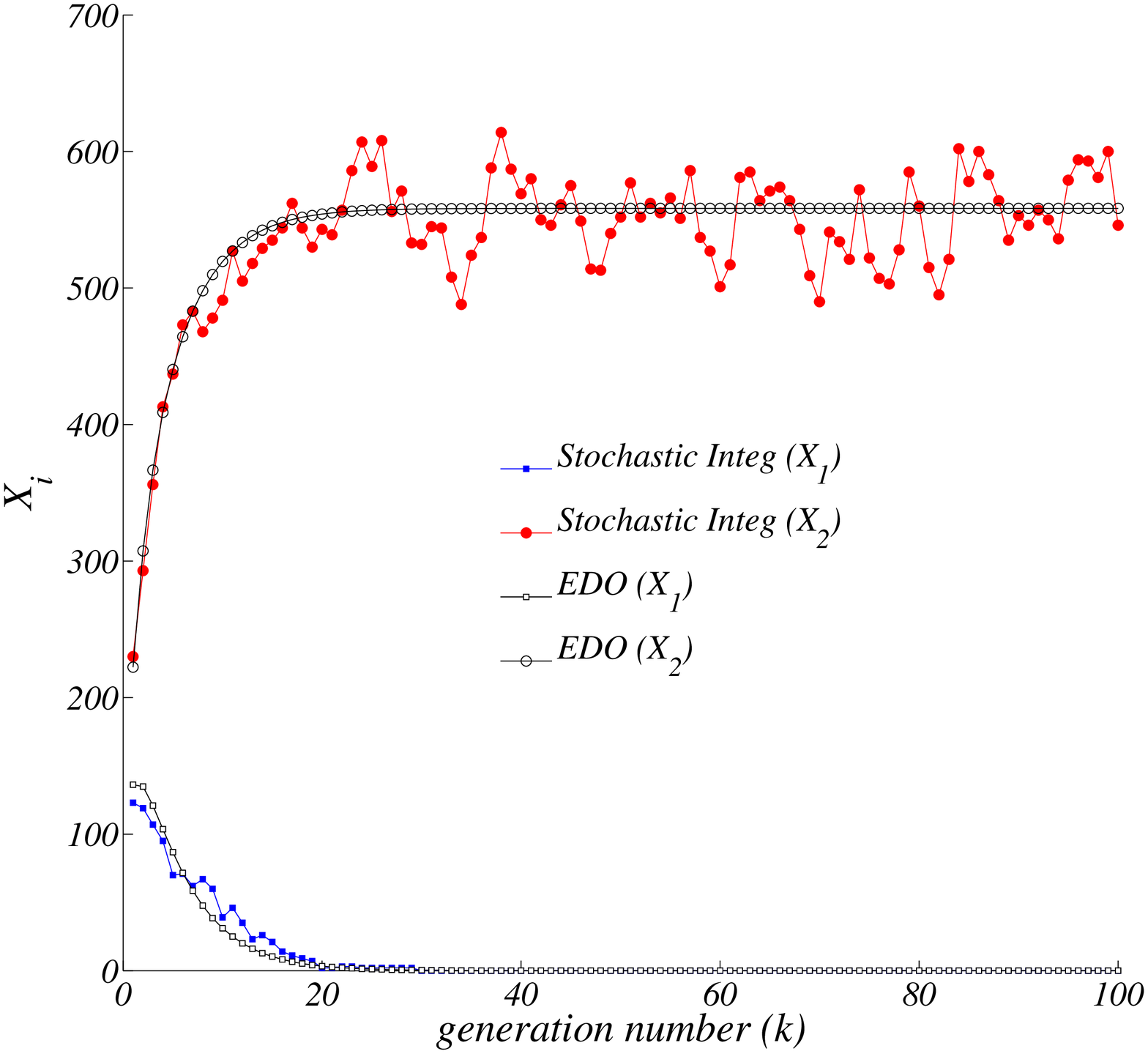}\quad \includegraphics[scale=0.2]{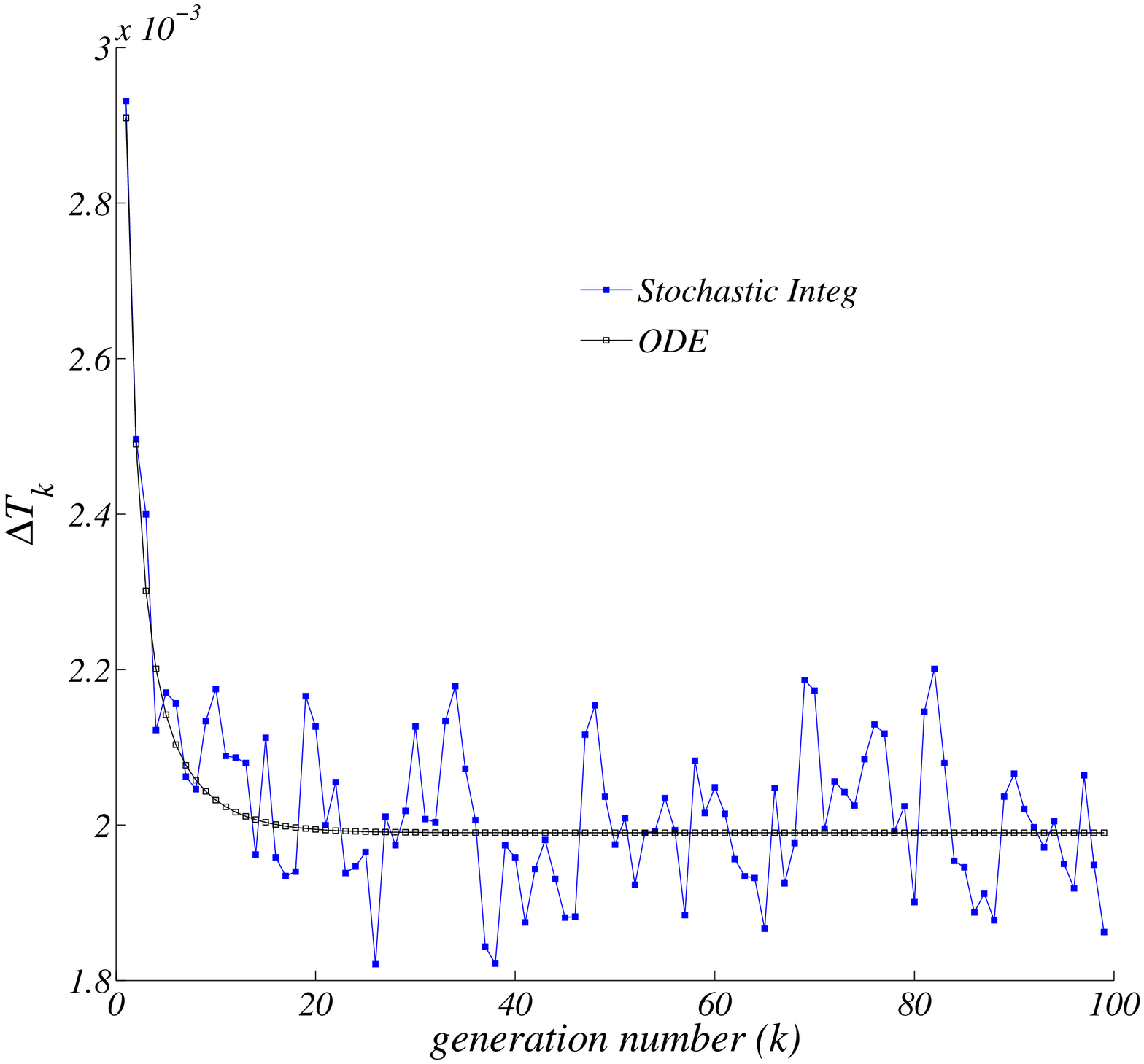} 
  \caption{Stochastic vs ODE SRM protocell~\eqref{eq:ode}. Case of two GMMs, left panel the
    amount of GMM at the beginning of each division cycle, right panel the
    division time as a function of the number of elapsed divisions. Parameters
    are : $\eta_1=\eta_2=1$, $\alpha_1=\alpha_2=2$, $L1=500$, $L2=600$,
    $P1=600$, $P_2=670$, $X_1(0)=X_2(0)=100$, $C(0)=1000$, $\rho=200$ and
    $\beta=2/3$.} 
  \label{fig:XYCSRN1divlarge}
  \end{center}
 \end{figure}
 \end{center}

As previously done, we compare the stochastic and the deterministic
models. Results are reported in Figure~\ref{fig:XYCSRM1divlarge} and one can
still observe that in presence of a large number of molecules the
deterministic dynamics well approximates the stochastic model. 
On the other hand, the protocell division time exhibits large fluctuations
around the deterministic value even in presence of a quite large number of
molecules (see right panel Fig.~\ref{fig:XYCSRN1divlarge}).

The parameters have been set in such a way only one GMM will survive according
to the analytical theory for the deterministic model. One can observe that,
despite the fluctuations, the same fate is obtained for the stochastic model
(see right panel Fig.~\ref{fig:XYCSRN1divlarge}).

\begin{center}
 \begin{figure}[ht]
  \begin{center}
  \includegraphics[scale=0.2]{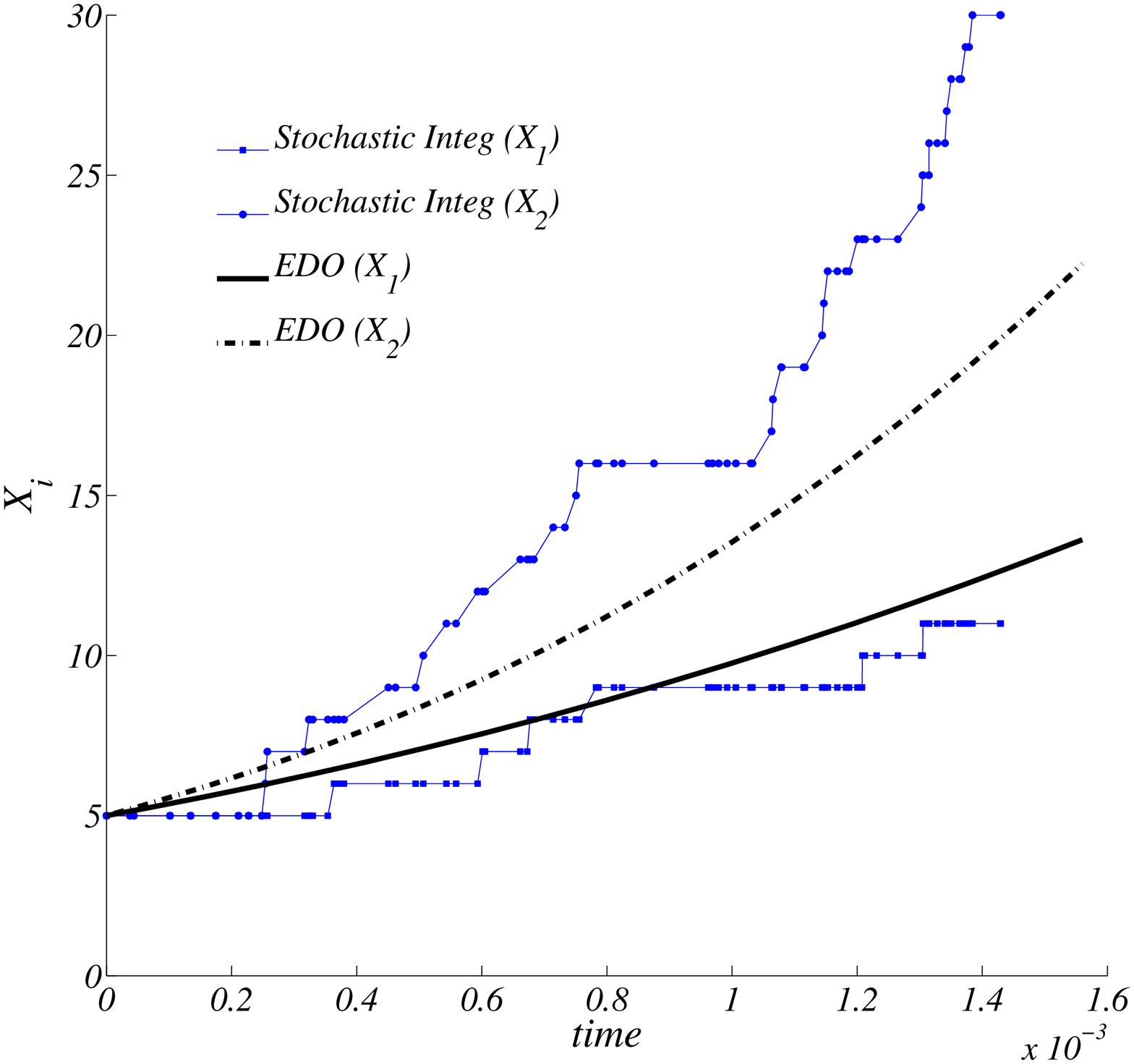}\quad \includegraphics[scale=0.2]{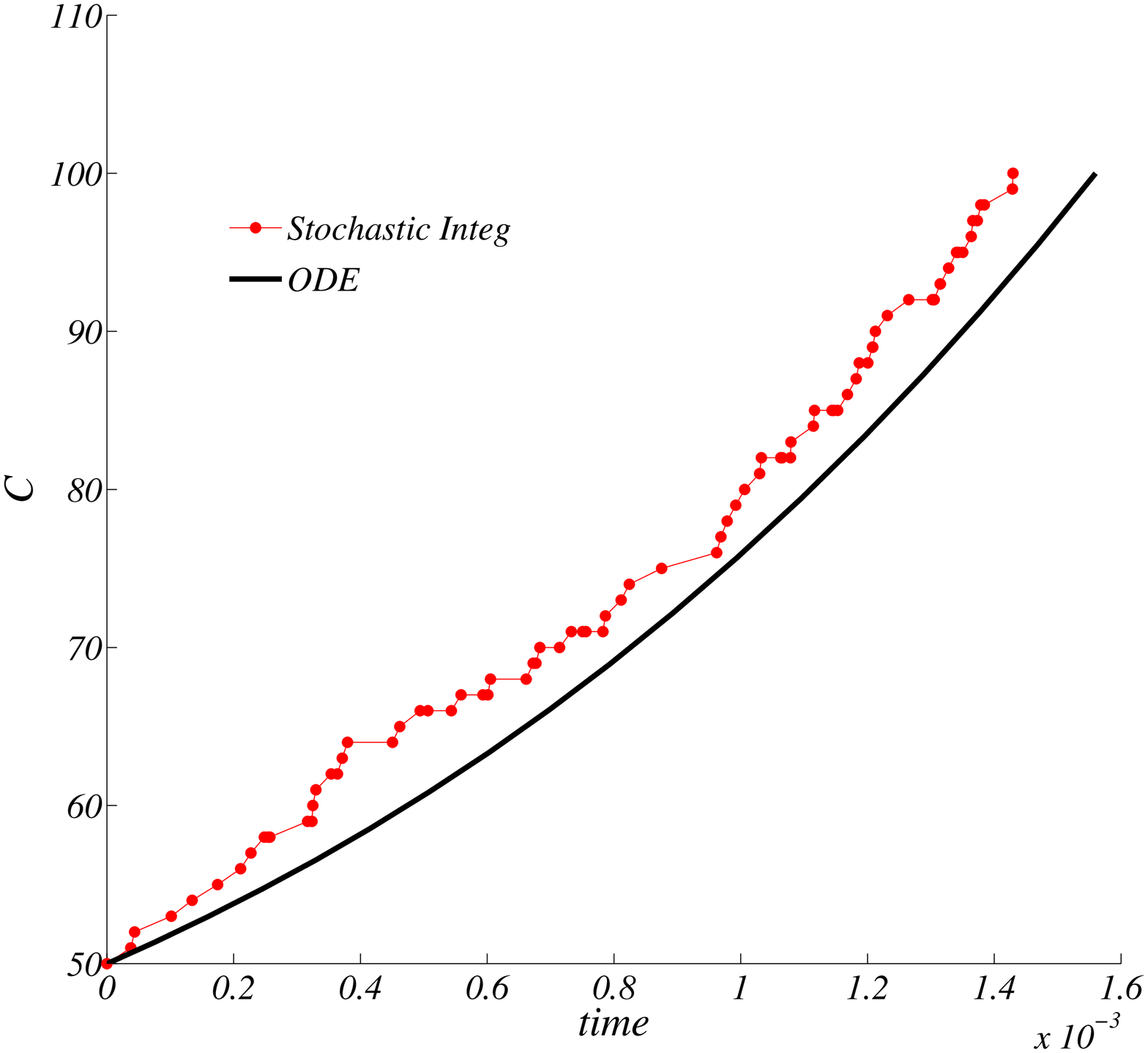} 
  \caption{Stochastic vs ODE SRM protocell~\eqref{eq:ode}. Case of two GMMs, left panel the
    time evolution of the amount of GMM during a division cycle, right panel
    the time evolution of the amount of $C$ molecules. Parameters
    are : $\eta_1=\eta_2=1$, $\alpha_1=\alpha_2=2$, $L1=500$, $L2=600$,
    $P1=450$, $P_2=670$, $X_1(0)=X_2(0)=5$, $C(0)=50$, $\rho=200$ and $\beta=2/3$.} 
  \label{fig:XCSRM1divlarge2}
  \end{center}
 \end{figure}
 \end{center}

\begin{center}
 \begin{figure}[ht]
  \begin{center}
  \includegraphics[scale=0.2]{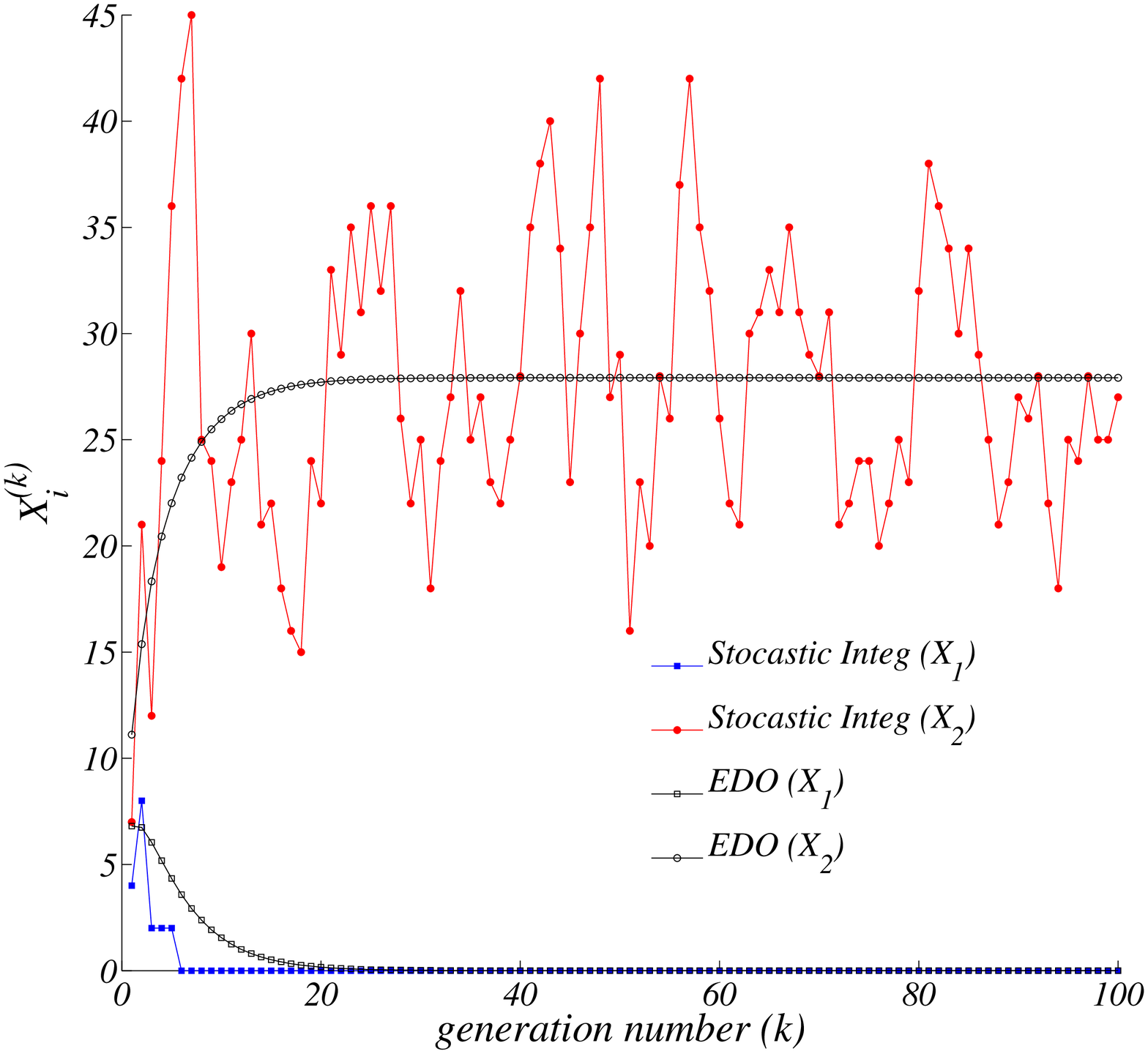}\quad \includegraphics[scale=0.2]{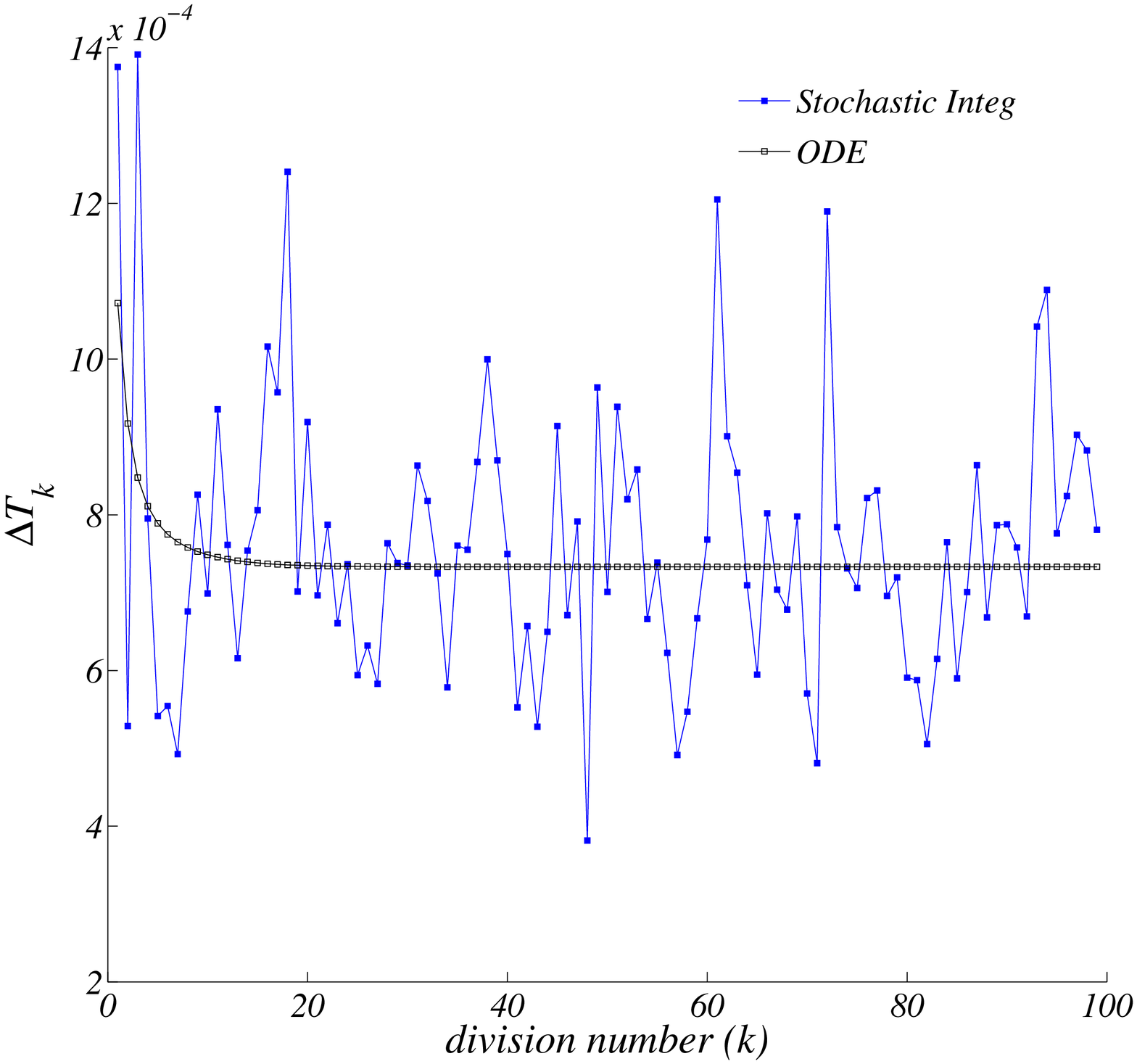} 
  \caption{Stochastic vs ODE SRM protocell~\eqref{eq:ode}. Case of two GMMs, left panel the
    amount of GMM at the beginning of each division cycle, right panel the
    division time as a function of the number of elapsed divisions. Parameters
    are : $\eta_1=\eta_2=1$, $\alpha_1=\alpha_2=2$, $L1=500$, $L2=600$,
    $P1=450$, $P_2=670$, $X_1(0)=X_2(0)=5$, $C(0)=50$, $\rho=200$ and
    $\beta=2/3$.} 
  \label{fig:XCSRMNdivlarge2}
  \end{center}
 \end{figure}
 \end{center}

Once we reduce the number of involved molecules, the stochastic fluctuations
dramatically increase (see Fig.~\ref{fig:XCSRM1divlarge2} and Fig.~\ref{fig:XCSRMNdivlarge2}).

\begin{center}
 \begin{figure}[ht]
  \begin{center}
  \includegraphics[scale=0.30]{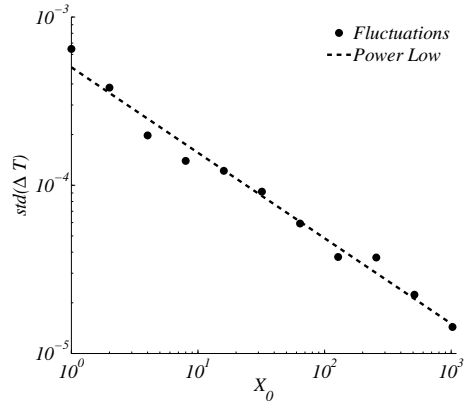}
  \caption{Fluctuation dependence on the initial conditions. We report the
standard deviation of the protocell division time as a function of the initial
amount of molecules $X_i(0)$, $i=1,2$, ($\bullet$) and a linear best fit,
whose slope is $=-0.51\pm 0.05$. Parameters are:
$X_{1}(0)=X_{2}(0)=2^{n}$ with $n=0,...,10$,~$C(0)=10X_{1}(0)$,~$\eta_1=\eta_2=1$,
$\alpha_1=\alpha_2=2$,~$L1=500$,~$L2=500$,~$P1=500$,~$P_2=600$,~$\rho=100$ and
$\beta=1$.} 
  \label{fig:XYCSRMSTDDeltaT}
  \end{center}
 \end{figure}
 \end{center}

As in the case of only one GMM, when two non interacting linear GMMs are
present the size of the stochastic fluctuations as a function of the initial
number of molecules follows a power law distribution with exponent $-0.51\pm 0.05$ (linear best fit), see Fig.~\ref{fig:XYCSRMSTDDeltaT}: the fewer are the molecules in the system, the
larger are the fluctuations around the deterministic dynamics.
 
\begin{center}
 \begin{figure}[ht]
  \begin{center}
  \includegraphics[scale=0.30]{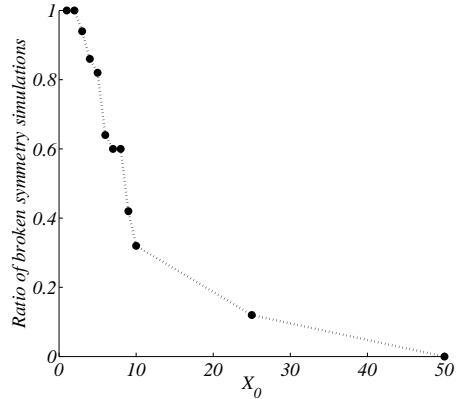}
  \caption{Symmetry breaking phenomenon. Each point denotes the fraction of runs exhibiting the symmetry breaking phenomenon, during $100$ generations, over $50$ identical replicas. Parameters are:  $X_{1}(0)=X_{2}(0)=[1,2,3,4,5,6,7,8,9,10,25,50]$,~$C(0)=10X$,~$\eta_1=\eta_2=1$, 
~$\alpha_1=\alpha_2=2$,~$L1=500$,~$L2=500$,~$P1=600$,~$P_2=600$,~$\rho=100$
and $\beta=1$.}
  \label{fig:XYCSRMsymmBroken}
  \end{center}
 \end{figure}
 \end{center}

A new phenomenon arises in the case of two GMMs modeled by a stochastic
process. There can be a {\em breaking of the symmetry} emerging in systems
composed of two identical GMMs (i.e equal kinetic 
  constants, equal initial amounts and availability of precursors) present
  with a few initial amount of each one. Although adopting a deterministic
  approach the dynamics 
  of the two replicators would be perfectly the same, a small fluctuation in
  the very first instants of the protocell evolution entails the dilution of
  one of the two replicators and thus a different fate for the
  protocell. Let us observe that the probability to have a large fluctuation
  is never zero, thus waiting {for} a sufficiently long time, a specie can always
  disappear from the system and thus giving rise to the the breaking of the
  symmetry phenomenon. See Fig.~\ref{fig:XYCSRMsymmBroken} where we report, as a
  function of the initial amount of molecules $X_i(0)$, $i=1,2$, the
  proportion of simulations where the symmetry breaking has been observed
  repeating $50$ times each simulation with the same set of parameters and
  initial conditions during $100$ generations.

\section{Conclusion}
\label{sec:concl}

In this paper we presented a new stochastic integration algorithm based on the
one introduced by Gillespie. Our contribution is devoted to the explicit
introduction of the volume variation in the algorithm, which moreover is
directly related to the amount of contained molecules, and thus it evolves in
a self-consistent way.

This algorithm straightforwardly adapts to the study of the evolution of a
protocell, simplified form of cells, where an ensemble of chemical reactions
occurs in a varying volume, the volume of the protocell, that in turn
increases because of the production of container molecules.

We presented several protocell models and we compare them with the analogous
deterministic protocell models, namely solved using the ODE. In this
preliminary study, we emphasized the role of the fluctuations and their
dependence on the initial amount of molecules. The dynamics is richer than the
deterministic one and thus it is worth studying, in particular we deserve to
future investigations the case where several molecules interact in a linear
way but including cross catalysis, i.e. the interaction matrix is not
diagonal, or they interact in a non-linear way. Also the study of the emergence
of time-periodic patterns due to 
the fluctuations, will be analyzed. An analytical treatment of the latter case
could be possible 
using some recent technics developed by~\cite{mckanenewman,dauxoisetal}, see
also~\cite{Anna2010} where the space is also taken into account.

% \subsection{Two interacting Genetic Memory Molecules : only catalysis}
% \label{ssec:2catggm}

%  \begin{center}
%  \begin{figure}[ht]
%   \begin{center}
%   \includegraphics[scale=0.2]{X1X2k_21042010_001.eps}\quad 
%   \includegraphics[scale=0.2]{DeltaT_21042010_001.eps} 
% %  \caption{protocell.$M=\left(\begin{smallmatrix}1 & 0.3 \\ 0.5 & 1 \end{smallmatrix}\right)$,$\alpha=(2,2)$, $L1=500$, $L2=600$,$P=\left(\begin{smallmatrix}500 & 530 \\ 500 & 540 \end{smallmatrix}\right)$,$X_1(0)=X_2(0)=100$, $C(0)=1000$, $\rho=200$, binomial divi,adiabatic assumption $\beta=2/3$.} 
% 	\caption{questa prova}
%   \label{fig:X1X2CSRMNdivlarge}
%   \end{center}
%  \end{figure}
%  \end{center}

%  \begin{center}
%  \begin{figure}[ht]
%   \begin{center}
%   \includegraphics[scale=0.2]{X1X2k_21042010_002.eps}\quad \includegraphics[scale=0.2]{DeltaT_21042010_002.eps} 
% %  \caption{protocell. $M=\left(\begin{smallmatrix}1 & 0.3 \\ 0.5 & 1 \end{smallmatrix}\right)$ $\alpha=(2,2)$ $L1=500$, $L2=600$ $P=\left(\begin{smallmatrix}500 & 530 \\ 500 & 540 \end{smallmatrix}\right)$ $X_1(0)=X_2(0)=5$ $C(0)=50$ $\rho=200$ binomial divi, adiabatic assumption $\beta=2/3$.}
% \caption{altra prvoa}
%   \label{fig:X1X2CSRMNdivsmall}
%   \end{center}
%  \end{figure}
%  \end{center}

% \subsection{Two interacting Genetic Memory Molecules : also inhibition}
% \label{ssec:2inibggm}

% TO BE DONE

\noindent
{\bf Acknowledgment}
The work of TC has been partially supported by the FNRS grant \lq\lq Mission
Scientifique 2010-2011\rq\rq. TC would like also to thank the European Center
for Living Technology in Venice (Italy) for the warm hospitality during a
short visiting stay that originates this collaboration.

\end{document}